\documentclass[twocolumn]{aa} % for a paper on 1 column  
\usepackage{txfonts}
\usepackage{graphicx,epsf}% Include figure files
\usepackage{amsmath,graphicx,adjustbox,setspace, sidecap, float}
\usepackage[toc,page]{appendix}
\usepackage{gensymb}
\usepackage{hyperref}
\usepackage{fancyhdr}
\usepackage{natbib}
\usepackage{caption,subcaption}
\usepackage{color}

\newcommand{\fg}[1]{Fig.~\ref{fig:#1}}
\newcommand{\Fg}[1]{Figure~\ref{fig:#1}}%beginning of the sentence

\newcommand{\Fgs}[2]{Figures\ \ref{fig:#1} and \ref{fig:#2}}

\newcommand{\eq}[1]{Eq.~(\ref{eq:#1})\xspace}
\newcommand{\Eq}[1]{Equation~(\ref{eq:#1})\xspace}%beginning of the sentence

\begin{document}

   \title{Forming Giant Planets Around Late-M Dwarfs: Pebble Accretion and Planet-Planet Collision}

   %\subtitle{I. Overviewing the $\kappa$-mechanism}

   \author{Mengrui Pan\inst{1} \and Beibei Liu\inst{1,\star} \and Anders Johansen\inst{2,3} \and Masahiro Ogihara\inst{4} \and Su Wang\inst{5,6} \and Jianghui Ji\inst{5,6,7} \and Sharon X. Wang\inst{8} \and Fabo Feng\inst{4} \and Ignasi Ribas\inst{9,10} 
          }

   \institute{Institute for Astronomy, School of Physics, Zhejiang University, Hangzhou 310027,  China\\
              \email{[panmr; bbliu]@zju.edu.cn}
              \and
               Center for Star and Planet Formation, GLOBE Institute, University of Copenhagen, {\O}ster Voldgade $5{-}7$, 1350 Copenhagen, Denmark
               %\email{anders.johansen@sund.ku.dk}
               \and
               Lund Observatory, Department of Astronomy and Theoretical Physics, Lund University, Box 43, 221 00 Lund, Sweden
               \and
               Tsung-Dao Lee Institute, Shanghai Jiao Tong University, 520 Shengrong Road, Shanghai 201210, China
               %\email{Ogihara@sjtu.edu.cn}
               \and
               CAS Key Laboratory of Planetary Sciences, Purple Mountain Observatory, Chinese Academy of Sciences, Nanjing 210023, China
               \and
               CAS Center for Excellence in Comparative Planetology, Hefei 230026, China
               \and
               School of Astronomy and Space Science, University of Science and Technology of China, Hefei 230026, China
               \and
               Department of Astronomy, Tsinghua University, Beijing 100084, People’s Republic of China
               \and
               %\email{[wangsu; jijh]@pmo.ac.cn}
               Institut de Ci{\`e}ncies de l'Espai (ICE, CSIC), Campus UAB, c/ Can Magrans s/n, 08193 Bellaterra, Barcelona, Spain
               \and
               Institut d'Estudis Espacials de Catalunya (IEEC), c/ Gran Capit{\`a} 2–4, 08034 Barcelona, Spain
             }

   \date{}

% \abstract{}{}{}{}{} 
% 5 {} token are mandatory

  \abstract
{We propose a pebble-driven core accretion scenario to explain the formation of giant planets around the late-M dwarfs of $M_{\star}{=}0.1{-}0.2 \ M_{\odot}$.  In order to explore the optimal disk conditions for giant planet, we perform N-body simulations to investigate the growth and dynamical evolution of both single and multiple protoplanets in the disks with both inner viscously heated and outer stellar irradiated regions. The initial masses of the protoplanets are either assumed to be equal to $0.01 \ M_{\oplus}$ or calculated based on the formula derived from streaming instability simulations.
Our findings indicate that massive planets are more likely to form in disks with longer lifetimes, higher solid masses, moderate to high levels of disk turbulence, and larger initial masses of protoplanets. In the single protoplanet growth cases, the highest planet core mass that can be reached is generally lower than the threshold necessary to trigger rapid gas accretion, which impedes the formation of giant planets.  
Nonetheless, in multi-protoplanet cases, the cores can exceed the pebble isolation mass barrier aided by frequent planet-planet collisions. This consequently speeds up their gas accretion and promotes giant planet formation, making the optimal parameter space to grow giant planets substantially wider. Taken together, our results suggest that even around very low-mass stellar hosts, the giant planets with orbital periods of ${\lesssim}100$ days are still likely to form when lunar-mass protoplanets first emerge from planetesimal accretion and then grow rapidly by a combination of pebble accretion and planet-planet collisions in disks with a high supply of pebble reservoir ${>}50 \ M_{\oplus}$ and turbulent level of $\alpha_{\rm t} {\sim} 10^{-3}{-}10^{-2}$. }


   \maketitle
%
%-------------------------------------------------------------------

\section{Introduction}
\label{sec:intro}

Late M dwarfs are the end tail of low-mass stars with a typical stellar mass of ${\approx}0.1{-}0.2 \ M_{\odot}$. Gas-dominated giant planets, on the other hand, represent the upper branch of the planetary populations. The giant planet systems around late M dwarfs serve as a useful benchmark for understanding the planet assembling processes, thereby providing important constraints on planet formation models under extreme environments. 
Studying how giant planets form around late M-dwarfs is thus a matter of great interest.

The occurrence rate of giant planets ($\eta_{\rm J}$) with a mass higher than $30 \ M_{\oplus}$ has been observed to correlate with stellar mass \citep{Johnson2010}. Exoplanet surveys have found that $\eta_{\rm J}$ around early M dwarfs is approximately $3{-}5\%$ for cold giant planets \citep{Johnson2010, Bonfils2013, Suzuki2016, Sabotta2021} and $0.27\%$ for hot Jupiters \citep{Gan2023b}. These values are even lower around mid-to-late dwarfs. For instance, \cite{Pass2023} indicated an upper occurrence rate of $1.5\%$ for planets more massive than Jupiter at the locations out to the water-ice line. \cite{Bryant2023} inferred the hot Jupiter occurrence rate of $0.14 \%$ around stars less massive than $0.26 \ M_{\odot}$. 

Despite the intrinsically low $\eta_{\rm J}$, a few such systems have been discovered around late dwarfs. For instance, the Transiting Exoplanet Survey Satellite (TESS) has confirmed one young, warm giant planet TOI-1227 b with a maximum mass of $M_{\rm p}{\approx}0.5 \ M_{\rm J}$ orbiting around a very-low-mass star of 0.17 $M_{\odot}$ \citep{Mann2022}. In addition, three gas giants have been detected by radial velocity surveys: GJ 3512 b and c \citep{Morales2019, Lopez2020, Ribas2023} and GJ 9066 c \citep{Feng2020, Quirrenbach2022}, all of which have masses comparable to that of Saturn. 
Meanwhile, microlensing observations contribute a high number of cold, massive planets/brown dwarfs around these stellar objects. However, the planet's masses and orbital properties are not well-constrained in most circumstances \citep{Suzuki2016, Zang2023}.

The lack of giant planets around low-mass stars can be naturally attributed to the scarcity of solid material in their protoplanetary disks. Observations of nearby star-forming regions at (sub)millimeter wavelengths have shown a steeper-than-linear correlation between the solid disk masses and stellar masses \citep{Pascucci2016, Ansdell2017}, although with a huge intrinsic scatter \citep{Manara2023}. This correlation implies a significant shortage of building blocks for planet formation around very low-mass stars. To form giant planets around such systems, an extremely high conversion efficiency from dust to planets would be required.

Several mechanisms have been proposed for giant planet formation.
The disk instability theory suggests that giant planets form directly through the gravitational fragmentation of young, massive protoplanetary disks \citep{Boss1997, Stamatellos2009, Deng2021, Boss2023}.
\cite{Mercer2020} explored the conditions for giant planet formation around stars with masses of $0.2{-}0.4 \ M_{\odot}$, and found that a super-massive disk (${>}30 \%$ of the host mass) is necessary to yield disk fragments, which typical results in planets with masses several times that of Jupiter at a few tens of au.  \cite{Morales2019} reported that gas giant GJ 3512 b could form through disk instability, with a similar disk mass requirement. However, in contrast to observations indicating an increase in $\eta_{\rm J}$  with higher stellar metallicity \citep{Santos2004, Fischer2005, Gan2022, Gan2023a}, the giant planets formed through gravitational instability appear to be largely unaffected by the disk metallicity \citep{Boss2002, Cai2006, Mercer2020}.

The core accretion theory suggests that the formation of giant planets requires the growth of massive cores (${ \sim }10 \ M_{\oplus}$) to initiate rapid gas accretion \citep{Pollack1996} before the dissipation of disk gas. Two channels for core accretion have been proposed: planetesimal accretion and pebble accretion. In the classical planetesimal-driven core accretion scenario \citep{Ida2004, Ogihara2009, Zhang2009, Mordasini2012, Coleman2016}, protoplanet growth occurs by accreting surrounding planetesimals with characteristic sizes of tens to hundreds of kilometers. \cite{Miguel2020} found that only Earth-analog systems could form around stars of $0.1{-0.2}\ M_{\odot}$ with sufficiently massive disks.  \cite{Burn2021} used sub-kilometer-sized planetesimals and explored the formation of giant planets by adjusting the initial planetesimal surface density and the type I migration rate. They concluded that giant planets can only form in massive disks (gas mass ${>}0.007 \ M_{\odot}$, solid mass ${>} 66 \ M_{\oplus}$) and when the type I migration rate is reduced by an order of magnitude (see, e.g., \cite{Ogihara2015, Ogihara2018}).
A follow-up study from \cite{Schlecker2022} indicated that with the current planetesimal accretion model, it is difficult to reproduce the observed giant planet population around stars less massive than $0.5 \ M_{\odot}$.

In the pebble-driven core accretion scenario \citep{Ormel2010, Lambrechts2012, Lambrechts2014a, Ida2016, Ormel2017a,Liu2019a, Liu2020b,Venturini2020, Chachan2023}, planets accrete millimeter-to-centimeter-sized pebbles to increase their masses. Compared to planetesimal accretion, the accretion cross-section of pebbles is largely enhanced by aerodynamic gas drag \citep{Ormel2017a, Johansen2017, Liu2020a}. \cite{Ormel2017b} and \cite{Schoonenberg2019} proposed a pebble-driven planet formation model for the TRAPPIST-1 system, and the resulting low water content and characteristic masses of all seven planets are in good agreement with observations. \cite{Coleman2019} examined both planetesimal and pebble accretion modes for this peculiar system and noted that diverse outcomes may arise by considering pebble ablation and planet atmosphere recycling.
Notably, in the context of pebble accretion scenarios, the core masses of the planets are limited by the pebble isolation mass \citep{Lambrechts2014b}, which is defined as the point when the planets reach a mass that opens a shallow gap in the protoplanetary disk and truncates the drifting pebbles. \cite{Liu2019a, Liu2020b} showed that planets around $0.1 
 \ M_{\odot}$ can only reach a maximum mass of $2{-}3 \ M_{\oplus}$ due to the low pebble isolation mass around these stellar hosts. Moreover, the characteristic masses of forming planets clearly exhibit a dependence on stellar metallicity \citep{Liu2019a}.

On the other hand, planets approaching such isolation masses can undergo substantial orbital migration. The strength and direction of migration are determined by the disk properties \citep{Paardekooper2011}. Planets at different disk radii can migrate convergently towards and become trapped in mean motion resonances \citep{Wang2017, Pan2022} at some special disk locations with zero net torque (also known as the transition radius, see \cite{Lyra2010, Horn2012, Kretke2012, Liu2015}). Such planet migration-induced mass concentration is likely to trigger dynamical instability with frequent orbital crossings and close encounters \citep{Zhang2014}. Massive cores can be attained through mutual planet-planet collisions in the gas-rich disk phase, promoting subsequent rapid gas accretion.  This planet-planet collision driven core accretion scenario has been mainly explored around solar-type stars \citep{Liu2015, Wimarsson2020} within a limited stellar mass range \citep{Liu2016}. Noticeably, inferred from the Juno measurement, Jupiter features a dilute core with an extended heavy element layer \citep{Wahl2017, Helled2017}. This pattern could be explained by the giant impacts among protoplanets during their final assembling phase \citep{LiuSF2019}.

In light of the literature studies presented, we speculate that the rapid growth of planetary cores can be achieved by a combination of the above two growth models. In this regard, we propose a hybrid growth model that utilizes both pebble accretion and planet-planet collisions to explain the formation of giant planets around late dwarfs. In this new scenario, the pebble isolation mass is not a barrier that limits core accretion, unlike in previous single protoplanet growth models \citep{Liu2019a, Liu2020b}. Pebble accretion plays a key role in the growth of individual protoplanets. As these protoplanets reach certain masses and migrate towards the transition radius, planet-planet collisions take over to assemble massive cores that can transition to runaway gas accretion. We evaluate the above hypothesis in this paper.

The paper is structured as follows. The model setup and N-body implementation are described in Section \ref{sec:method}. The growth of individual protoplanets with different disk and planet parameters is examined in Section \ref{sec:single}. Section \ref{sec:multi-protoplanets} explores the formation and evolution of multiple protoplanets, and Section \ref{sec:discussion}  presents an assessment of the model and its implications. Finally, we summarize the key results in Section \ref{sec:conclusion}.

\section{Method}
\label{sec:method}

We adopt the planet formation model from \cite{Liu2019a}.
We summarize the key physical processes and main equations here. A detailed and complete model description is referred to Section 2 of \cite{Liu2019a}.

\subsection{Disk model}
We employ the $1$D standard viscous $\alpha-$disk model \cite{Shakura1973} and assume the disk evolves in a quasi-steady manner. The disk is divided into two distinct components based on different heating mechanisms \citep{Garaud2007}. The inner optically thick disk region is viscously heated \citep{Ruden1986}, in which the gas surface density, temperature and disk aspect ratio are given by 
\begin{equation}
    \begin{aligned}
        \Sigma_{\rm g,vis} = \ & 99 \left( \frac{\dot{M}_{\rm g}}{10^{-8} \ {\rm M}_{\odot} \ {\rm yr}^{-1}} \right)^{1/2} \left( \frac{M_{\star}}{0.1 \ {\rm M}_{\odot}} \right)^{1/8} \left( \frac{\alpha_g}{10^{-2}} \right)^{-3/4} \\
        & \times \left( \frac{\kappa_0}{10^{-2}} \right)^{-1/4} \left( \frac{r}{1 \ \rm au} \right)^{-3/8} \rm g \ cm^{-2},
    \end{aligned}   
\end{equation}
\begin{equation}
    \begin{aligned}
        T_{\rm g,vis} = \ & 118 \left( \frac{\dot{M}_{\rm g}}{10^{-8} \ {\rm M}_{\odot} \ {\rm yr}^{-1}} \right)^{1/2} \left( \frac{M_{\star}}{0.1 \ {\rm M}_{\odot}} \right)^{3/8} \left( \frac{\alpha_g}{10^{-2}} \right)^{-1/4} \\
        & \times \left( \frac{\kappa_0}{10^{-2}} \right)^{1/4} \left( \frac{r}{1 \ \rm au} \right)^{-9/8} \rm K,
    \end{aligned}   
\end{equation}
\begin{equation}
    \begin{aligned}
        h_{\rm g,vis} = \ & 0.07 \left( \frac{\dot{M}_{\rm g}}{10^{-8} \ {\rm M}_{\odot} \ {\rm yr}^{-1}} \right)^{1/4} \left( \frac{M_{\star}}{0.1 \ {\rm M}_{\odot}} \right)^{-5/16} \left( \frac{\alpha_g}{10^{-2}} \right)^{-1/8} \\
        & \times \left( \frac{\kappa_0}{10^{-2}} \right)^{1/8} \left( \frac{r}{1 \ \rm au} \right)^{-1/16},
        \label{eq:hvis}
    \end{aligned}   
\end{equation}
and $\dot{M}_{\rm g}$, $M_{\star}$, and $r$ are the disk accretion rate, stellar mass, and the distance to the central star, respectively. At 1 au, the initial gas surface density is approximately $100 \ {\rm g} \, {\rm cm}^{-2}$. Given the low surface densities, the magneto-rotational instability (MRI) might be driven by cosmic-ray ionization\cite{Gammie1996}. In this paper, we focus on late M dwarfs; therefore, the stellar mass is specified as $M_{\star}{=}0.1 \ M_{\odot}$ unless otherwise explored (see Section \ref{sec:starmass}). We assume that the disk opacity is $\kappa{=} \kappa_0 \left( T_{\rm g}/1 \ {\rm K} \right) \ {\rm g} \, {\rm cm}^{-2}$ \citep{Garaud2007}, where $\kappa_0=0.01$ is the opacity coefficient, and $\alpha_{\rm g}$ represents the global disk angular momentum transport efficiency \cite{Shakura1973}. We choose $\alpha_{\rm g}{=}10^{-2}$, inferred from disk observations of $\dot M_{\rm g}$ and $\Sigma_{\rm g}$ \citep{Hartmann1998, Andrews2009}.

The outer disk is assumed to be optically thin in the vertical direction and primarily heated by stellar irradiation \citep{Chiang1997}, in which the gas surface density, temperature and disk aspect ratio are given by \citep{Ida2016}
\begin{equation}
    \begin{aligned}
        \Sigma_{\rm g,irr} = \ & 212 \left( \frac{\dot{M}_{\rm g}}{10^{-8} \ {\rm M}_{\odot} \ {\rm yr}^{-1}} \right) \left( \frac{M_{\star}}{0.1 \ {\rm M}_{\odot}} \right)^{9/14} \left( \frac{L_{\star}}{0.01 \ {\rm L}_{\odot}} \right)^{-2/7} \\
        & \times\left( \frac{\alpha_g}{10^{-2}} \right)^{-1} \left( \frac{r}{1 \ \rm au} \right)^{-15/14} \rm g \ cm^{-2},
    \end{aligned}   
\end{equation}
\begin{equation}
    \begin{aligned}
        T_{\rm g,irr} = \ & 56 \left( \frac{M_{\star}}{0.1 \ {\rm M}_{\odot}} \right)^{-1/7} \left( \frac{L_{\star}}{0.01 \ {\rm L}_{\odot}} \right)^{2/7} \left( \frac{r}{1 \ \rm au} \right)^{-3/7} \rm K,
    \end{aligned}   
\end{equation}
\begin{equation}
    \begin{aligned}
        h_{\rm g,irr} = \ & 0.047 \left( \frac{M_{\star}}{0.1 \ {\rm M}_{\odot}} \right)^{-4/7} \left( \frac{L_{\star}}{0.01 \ {\rm L}_{\odot}} \right)^{1/7} \left( \frac{r}{1 \ \rm au} \right)^{2/7},
        \label{eq:hirr}
    \end{aligned}
\end{equation}
where $L_{\star}$ is the stellar luminosity. 
We use $f_s=1/(1+r_{\rm tran})^4$ as a smooth function to combine the inner and outer regions, therefore the global disk quantity can be calculated by $X = X_{\rm vis}f + (1-f)X_{\rm irr}$.

The transition radius between the inner and outer regions is given by
\begin{equation}
    \begin{aligned}
        r_{\rm tran} = \ & 3.0 \left( \frac{\dot{M}_{\rm g}}{10^{-8} \ {\rm M}_{\odot} \ {\rm yr}^{-1}} \right)^{28/39} \left( \frac{M_{\star}}{0.1 \ {\rm M}_{\odot}} \right)^{29/39} \\
        & \left( \frac{L_{\star}}{0.01 \ {\rm L}_{\odot}} \right)^{-16/39} \left( \frac{\alpha_g}{10^{-2}} \right)^{-14/39} \left( \frac{\kappa_0}{10^{-2}} \right)^{14/39} \ {\rm au},
    \end{aligned}   
\end{equation}
which decreases as disk dissipation. 

Initially the young disks can maintain a continuous supply of infall of material from their parent molecular clouds \citep{Padoan2014}. At later times the infall is quenched and the disks gradually deplete gas by combined effects of viscous accretion and stellar photoevaporation \citep{Hartmann1998, Alexander2014, Ercolano2018}.  In this work we simply assume that the disk accretion rate remains a constant $ \dot{M}_{\rm g} {=} \dot{M}_{\rm g, 0}$ at $t {\leq} t_0$, and follows an exponential decay $\dot{M}_{\rm g}{ =} \dot{M}_{\rm g, 0} \exp{[-(t-t_0)/\tau_{\rm dep}]}$ at $t{>}t_0$ where $t_0$ separates the early infall and later dissipation stages, and $\tau_{\rm dep}$ is the disk depletion timescale. The total gas disk mass is therefore parameterizedly described by $\dot{M}_{\rm g, 0}, t_0$ and $ \tau_{\rm dep}$.

Observations of disk accretion rate onto very low-mass stars show a wide spread, ranging from ${<}10^{-9} \ {\rm M}_{\odot} \ {\rm yr}^{-1}$ up to $1{-}2 \times 10^{-8} \ {\rm M}_{\odot} \ {\rm yr}^{-1}$ \citep{Hartmann2016, Pinilla2021}. In this paper we focus on giant planet formation around late dwarfs. The paucity of massive planets around these stars indicates such systems are expected to grow only in relatively massive disks. Hence, in the fiducial model for studying the systems around stars of $0.1 \ \rm  M_{\odot}$, we choose a relatively high initial disk accretion rate of $\dot{M}_{\rm g,0} {=} 10^{-8} \ {\rm M}_{\odot} \ {\rm yr}^{-1}$, and the disk starts to dissipate at $t_0 {=} 1$ Myr on a timescale $\tau_{\rm dep}$ of $0.5$ Myr. Here we define the disk lifetime $t_{\rm disk}$ as the time when $\Sigma_{\rm g}$ at 1 au drops below $1 \ \rm g \ cm^{-2}$.  In this circumstance, the initial disk mass is $15\%$ of its stellar mass and $t_{\rm disk}{=}3.7$ Myr.

On the other hand, the disk lifetime is inferred to be longer around lower-mass stars \citep{Williams2011, Bayo2012, Manara2012, Ribas2015, Picogna2021}. In order to investigate the influence of disk lifetime on planet formation, we construct a disk with the same initial mass as the fiducial one but vary $\dot{M}_{\rm g,0} {=} 6 \times 10^{-9} \ {\rm M}_{\odot} \ {\rm yr}^{-1}$, $t_0 {=} 1.5$ Myr and  $\tau_{\rm dep}=1$ Myr. In such a circumstance $t_{\rm disk}{=}6.3$ Myr.

\subsection{Growth and migration of protoplanet}
\subsubsection{Initial mass of protoplanet}

We start the growth of protoplanet with an initial mass $M_{\rm p0}$. 
There are two considerations regarding the choice of $M_{\rm p0}$.  First, a canonical value of $0.01 \ M_{\oplus}$ is widely adopted in literature, which can date back to the pioneering numerical N-body simulation study of \cite{Kokubo1998}. They found that a few lunar-mass oligarchs naturally emerge out from a swarm of small planetesimals by mutual collisions. Their study is limited to the circumstances of host stars with a solar mass. The formation of embryos with $0.01{-}0.1 \ M_{\oplus}$ around M dwarfs after oligarchic growth has also been suggested by \cite{Ogihara2009} (note that their study was conducted under the assumption of the solar nebular conditions). No further extended numerical work has been conducted to explore how the forming masses of protoplanets around lower-mass dwarfs. \cite{Ormel2010b} analytically derived the transition mass between runaway and oligarchic growth such that $M_0 {\propto} M_{\star}^{-3/7}\Sigma_{\rm plt}^{6/7} R_{\rm plt}^{9/7}$ (their Eq.13), where $\Sigma_{\rm plt}$ and $R_{\rm plt}$ are the surface density and size of planetesimals. The latter two quantities ($\Sigma_{\rm plt}$ and $R_{\rm plt}$) should be also dependent on the host stellar environment.  Therefore, without any further assumptions on the planetesimal formation models, the exact stellar mass dependency of $M_0$ is unknown. Bearing these uncertainties, we follow similar literature studies \citep{Liu2019a,Burn2021} and adopt $M_{\rm p0}{=}0.01 \ M_{\oplus}$  as a prior in the follow-up explorations. This constant mass assumption can serve as one benchmark, which differs from the secondary consideration that specifically assumed one planetesimal formation model.  We also note that planets with masses that exceed this value are well in the settling pebble accretion regime, where the planet-pebble interaction is substantially aided by gas drag \cite{Ormel2010, Liu2018, Liu2019b}.

On the other hand, streaming instability provides a valuable pathway for the formation of planetesimals \citep{Youdin2005, Johansen2007}.
It occurs when the volume density of pebbles approaches that of the gas, resulting in significant back-reaction from pebbles onto the gas. As a result, these pebbles concentrate radially into dense clumps and eventually collapse into planetesimals through self-gravity. Defining the protoplanet as the largest planetesimal generated from the streaming instability clumps, \cite{Liu2020b} obtain the mass of the protoplanet from the extrapolation of literature numerical investigations (e.g., \cite{Johansen2015, Simon2016, Schafer2017, Abod2019}). This streaming instability-induced protoplanet mass can be expressed as (see section 2.4 of \cite{Liu2020b} for derivations)

\begin{equation}
    \begin{aligned}
        M_{\rm p0} = & 2 \times 10^{-3} \left( \frac{\gamma}{\pi^{-1}} \right)^{3/2} \left( \frac{h_{\rm g}}{0.05} \right)^{3} \left( \frac{M_{\star}}{0.1 \ {\rm M}_{\odot}} \right) \ { M}_{\oplus}.
        \label{eq:Mp0}
    \end{aligned}
\end{equation}
Where $\gamma {=} 4\pi G \rho_{\rm g}/\Omega_{\rm K}^2$ is the relative strength between self-gravity and tidal shear, $\rho_{\rm g}{=}\Sigma_{\rm g}/(\sqrt{2 \pi }h_{\rm g} r)$ is the gas volume density and $\Omega_{\rm K}{=}\sqrt{GM_{\star}/r^3}$ is the Keplerian angular velocity.

To summarize,  we assume two scenarios for the starting mass of a protoplanet. In the equal-mass scenario, protoplanets form from classical planetesimal accretion \citep{Kokubo1998}. We assume they all have $M_{\rm p0}{=}0.01 \ {\rm M}_{\oplus}$. In the second scenario the protoplanets are specifically generated by streaming instability, and their birth masses follow \eq{Mp0}. 
In contrast to the equal-mass scenario and as can be seen in \eq{Mp0}, the mass of protoplanets formed by streaming instability correlates with gas disk density, gas disk aspect ratio and stellar mass. In this respect, we expect a higher $M_{\rm p0}$ at a larger orbital distance since both $\gamma$ and $h_{\rm g}$ increase with $r$.

\subsubsection{Pebble accretion} \label{subsubsec:PA}
Pebbles undergo fast radial drift towards the central star. A fraction of these drifting pebbles can be accreted by the planet when they cross the planetary orbit. The pebble accretion rate onto the planet's core is given by
 \begin{equation}
 \dot{M}_{\rm PA}  = \varepsilon_{\rm PA} \dot{M}_{\rm peb} = \varepsilon_{\rm PA} \xi_{\rm p/g}  \dot{M}_{\rm g} = \left(\varepsilon_{\rm PA,2D}^{-2} + \varepsilon_{\rm PA,3D}^{-2}\right)^{-1/2}  \xi_{\rm p/g}  \dot{M}_{\rm g}
\end{equation}
where $\dot{M}_{\rm peb}$ is the pebble mass flux and $\varepsilon_{\rm PA}$ is the total pebble accretion efficiency, the formulas of which are adopted from \cite{Liu2018} and \cite{Ormel2018} that include both 2D and 3D accretion efficiencies ($\varepsilon_{\rm PA,2D}$ and $\varepsilon_{\rm PA,3D}$) taking into account the eccentricity and inclination of the planet. In brief, the pebble accretion efficiency firstly gets boosted when the planets have relatively low eccentricity. It then drops with the further increase of eccentricity due to the fact that high pebble-planet impact is not in the settle regime anymore. On the other hand, the pebble accretion efficiency decreases with inclination since the planets are more likely to lift off the pebble plane when they are on inclined orbits.

In the limit of zero eccentricities and inclinations, the pebble accretion efficiency in the settling regime can be approximated as
\begin{equation}
        \varepsilon_{\rm PA,2D} = \frac{0.32}{\eta} \sqrt{ \frac{M_{\rm p}}{M_{\star}} \frac{1}{\tau_{\rm s}} \frac{\Delta v}{v_{\rm K}}}, \ \  
\varepsilon_{\rm PA,3D} = \frac{0.39}{\eta h_{\rm peb}} \frac{ M_{\rm p}}{ M_{\star}},
\end{equation}
where $v_{\rm K}$ is the Keplerian velocity,  $\Delta v$ is the relative velocity between the pebbles and planet (dominated by $\eta v_{\rm K}$ in the headwind regime, $\Omega_{\rm K} R_{\rm H}$ in the shear regime, and $R_{\rm H}{=}(M_{\rm p}/3M_{\star})^{3}r$ is the planet Hill radius),  $h_{\rm peb}$ is the pebble disk aspect ratio, $\eta = -h_{\rm g}^2 (\partial \ln P /\partial \ln r)/2$ and $P$ is the gas disk pressure. 
It is important to recognize that both $\varepsilon_{\rm PA,2D}$ and $\varepsilon_{\rm PA,3D}$ increase as $h_{\rm g}$ (or equivalently $\eta$) decreases. Physically, the pebble accretion efficiency becomes higher when the inward drifting pebbles is slower in the $2$D regime and/or the pebble disk is less vertically extended in the $3$D regime.

The pebble disk scale height  $H_{\rm peb}{ =} \sqrt{\alpha_{\rm t}/(\alpha_{\rm t}+\tau_{\rm s}}) \ H_{\rm g}$ \citep{Youdin2007},
where $\alpha_{\rm t}$ is the turbulent diffusion coefficient, approximately equivalent to the local turbulent viscous parameter when the disk is driven by magneto-rotational instability \citep{Johansen2005, Zhu2015}.
Physically, $\alpha_{\rm t}$ can differ from $\alpha_{\rm g}$ - the average value of the global disk angular momentum transport efficiency - due to instances of layered accretion \citep{Turner2008}. 
The midplane of the disk is quiescent and the high altitude region is turbulent active. We note that $\alpha_{\rm t}$ is more relevant to the midplane of the dead zone while $\alpha_{\rm g}$ represents the vertically average, global disk angular momentum transport efficiency. These two parameter are not always equal. The planet gap opening occurs at the disk midplane, and this process can also be closed by local turbulent diffusion.
Hence, $\alpha_{\rm t}$ is more relevant to the planet formation processes such as dust stirring, pebble accretion and gap opening \citep{Xu2017}.

Meanwhile, $\tau_{\rm s}$ is the pebble's dimensionless stopping time (termed Stokes number hereafter) that characterizes the aerodynamic size of pebbles. The detailed dust evolution is not modeled here. Advanced dust coagulation studies find that the largest pebbles dominate the total mass of the population and their Stokes number is almost a constant (e.g., in the fragmentation-limited regime).
For the sake of simplicity, we ideally treat that all pebbles reach a fixed Stokes number of $\tau_{\rm s}{ =}0.05$. The potential influence of  $\alpha_{\rm t}$ on $\tau_{\rm s}$ is discussed in Section \ref{sec:stokes}.

The pebbles are assumed to be constituted of $50\%$ water ice and $50\%$ silicate. The water-ice line $r_{\rm H_2O}$ is calculated when the disk temperature is $170$ K. When the pebbles drift inside of the water-ice line, their icy component sublimates, and the pebble mass flux decreases accordingly. We neglect the pebbles' Stokes number variation when they cross $r_{\rm H2O}$.

Same as \cite{Liu2019a, Liu2020b}, we assume that the pebble and gas flux ratio remains a constant such that $\xi_{\rm p/g}{=}\dot M_{\rm peb} / \dot M_{\rm g}$. Pebbles are well-coupled to the disk gas when their Stokes number is very low. Thus, pebbles and gas drift at the same speed and the initial disk metallicity is preserved, where disk metallicity $Z {=}\Sigma_{\rm peb}/ \Sigma_{\rm g}$. When the pebbles have a higher Stokes number, they drift faster than disk gas. In this case, in order to maintain a constant flux ratio, $\Sigma_{\rm peb}/ \Sigma_{\rm g}$ becomes lower than the initial disk metallicity. 
We assume $\xi_{\rm p/g} {=}0.01$ in the fiducial model, corresponding to totally  $50 \ M_{\oplus}$ solid in pebbles.   
It is worth noting that the above constant mass flux ratio is a global concept. The disk metallicity can still be enriched at local places due to various mechanisms. For instance, several studies proposed that the local solid density can be enhanced at the water-ice line \citep{Ros2013, Schoonenberg2017, Drazkowska2017}. We do not take these localized effects into account in this study.  The enrichment of disk metallicity in the late gas disk dispersal phase is also not considered.

As the planet grows, it becomes massive enough to perturb the surrounding gas and produce a local pressure bump. The inward drifting pebbles stop at the outer edge of the gap generated by the planet. As such, the planet cannot further accrete pebbles. This onset planet mass is defined as the pebble isolation mass \citep{Lambrechts2014b}. On the other hand, the gap opening mass is typically defined when the planet opens a gap whose surface density drops by $50\%$. We adopt the gap opening mass based on \cite{Kanagawa2015}'s 2D hydrodynamical simulations,  
\begin{equation}
    M_{\rm gap} = 5.8 \ \left(\frac{\alpha_{\rm t}}{10^{-3}}\right)^{1/2} \ \left(\frac{h_{\rm g}}{0.065}\right)^{5/2} \ \left(\frac{M_{\star}}{0.1 \ {\rm M}_{\odot}}\right) \ {\rm M}_{\oplus}.
\end{equation}

Based on the $1$D numerical simulations conducted by \cite{Johansen2019}, the pebble isolation mass is approximately $2.3$ times lower than the gap opening mass. We apply this scaling  and convert \cite{Kanagawa2015}'s gap opening mass into pebble isolation mass, which reads
\begin{equation}
    M_{\rm iso} = 2.5 \ \left(\frac{\alpha_{\rm t}}{10^{-3}}\right)^{1/2} \ \left(\frac{h_{\rm g}}{0.065}\right)^{5/2} \ \left(\frac{M_{\star}}{0.1 \ {\rm M}_{\odot}}\right) \ {\rm M}_{\oplus}.
    \label{eq:Miso}
\end{equation}
We also demonstrate a comparison of $M_{\rm iso}$ adopted in this work and other literature studies \citep{Ataiee2018,Bitsch2018} in Appendix \ref{app:iso}.

Same as \cite{Liu2019b} and \cite{Jang2022}, we consider the filtering of flux when pebbles drift through different planets in a multi-planetary system (see Eq.12 of \cite{Liu2019b}). That means the pebble mass flux entering the inner disk can be reduced due to the accretion of planets in the outer disk region. We simplified that the pebble accretion of the planets that reside in the interior of its orbit is terminated when the planet reaches $M_{\rm iso}$. The diffusion of small, fragmented particles through the gap is not considered \citep{Liu2022, Stammler2023}.

\subsubsection{Gas accretion}
Gas accretion can be divided into the early hydrostatic phase and later runaway phase \citep{Pollack1996}.
During hydrostatic accretion, the planet slowly captures disk gas to form a tiny atmosphere. The envelope hydrostatic equilibrium is established when the gravitational energy is balanced by radiative heating. The heat from solid accretion is quenched when the planet reaches the pebble isolation mass. The envelope is expected to undergo subsequent Kelvin-Helmholtz contraction.
For simplicity, we ignore gas accretion in the hydrostatic phase and treat $M_{\rm iso}$ as the onset mass for gas accretion\citep{Ogihara2020}. We note here that the literature critical core mass is estimated to be $5{-}15\ M_{\oplus}$ \citep{Ida2004, Alibert2019}, higher than $M_{\rm iso}$ (typically $1{-}2 \ M_{\oplus}$) around very low-mass stars.
Our simplification remains justified as the gravitational force of planets with isolation mass is insufficient to retain a substantial atmospheric envelope \citep{Alibert2019}.

The gas accretion rate in the Kelvin-Helmholtz contraction reads \citep{Ikoma2000} 
\begin{equation}
    \left( \frac{{\rm d} M_{\rm p, g}}{{\rm d} t} \right)_{\rm KH} = \ 8 \times 10^{-8} \ \left( \frac{M_{\rm p}}{3 \ {\rm M}_{\oplus}}\right)^{4} \ \left( \frac{\kappa_{\rm env}}{1 \ {\rm cm}^2 \ {\rm g}^{-1}}\right)^{-1} \ {\rm M}_{\oplus} \ {\rm yr}^{-1},
    \label{eq:KH}
\end{equation}
where $\kappa_{\rm env}$ is the envelope opacity, a crucial parameter that sets the amount of gas accreted by the planet. A variety of $\kappa_{\rm env}$ have been tested and it has been found that a very low $\kappa_{\rm env} {\ll} 0.1 \rm \ cm^{2}/g $ might lead overpopulated massive giant planets, contradicting with observations. On the other hand, the disk opacity is estimated to be ${\sim}1 \rm \  cm^2/g$ close to and beyond $r_{\rm H_2 O}$ based on an ISM-like dust size distribution \citep{Bell1994}.  The envelope opacity is expected to be no higher than the disk opacity. This is because when the planet reaches $M_{\rm iso}$,  large pebbles get completely blocked, whereas only small dust well coupled to the gas can drift across the gap and get accreted onto the planet \citep{Liu2022, Stammler2023}. This dust-size filtration lowers the opacity in the planet envelope compared to the disk gas. In addition, the envelope opacity could be further reduced by grain sedimentation, coagulation and evaporation \citep{Movshovitz2010, Ormel2014, Mordasini2014}. Considering the above reasons,  we adopt a moderate $\kappa_{\rm env}{=}0.1 \rm \ cm^{2}/g$ and assume it does not vary with the disk metallicity (see Fig.8 of \cite{Mordasini2014}).

The gas accretion in \Eq{KH}  decreases dramatically with the lowering of the planet mass. For instance, $ \dot M_{\rm p, g}{\sim} 1.5 \times 10^{-7} \  {\rm M}_{\oplus} \ {\rm yr}^{-1}$ at  $M_{\rm p}{=} 2 \ M_{\oplus}$, indicating the gas contraction is very limited over the disk lifetime in this circumstance. However,  $\dot M_{\rm p, g} {\sim} 4\times 10^{-5} \ {\rm M}_{\oplus} \ {\rm yr}^{-1}$ at $M_{\rm p}=8  \ M_{\oplus}$ and the planet double its mass within a few $10^{5}$ yr. The mass of planetary core plays a critical role in the gas accretion. It significantly affects the accretion rate and the total amount of gas that the planet accumulates before disk dissipation.

Besides, only a fraction of gas within the planet Hill sphere can be accreted \citep{Tanigawa2002, Machida2010}. We adopt the corresponding accretion rate from Eq. 29 of \citet{Liu2019a}:
\begin{equation}
    \begin{aligned}
        \left( \frac{{\rm d} M_{\rm p, g}}{{\rm d} t} \right)_{\rm Hill} = \ & 0.004  \left( \frac{M_{\rm p}}{3 \ {\rm M}_{\oplus}}\right)^{2/3} \ \left( \frac{M_{\star}}{0.1 \ M_{\odot}} \right)^{-2/3} \ \left( \frac{\dot{M}_{\rm g}}{10^{-8} \ {\rm M}_{\odot} \ {\rm yr}^{-1}}\right) \\ & \times \left( \frac{\alpha_{\rm g}}{10^{-2}}\right)^{-1} \ \left( \frac{h_{\rm g}}{0.065}\right)^{-2} \ \left[ 1 + \left( \frac{M_{\rm p}}{M_{\rm gap}} \right)^2 \right]^{-1} \ {\rm M}_{\oplus} \ {\rm yr}^{-1}.
    \end{aligned}
\end{equation}
The further gas accretion onto the planet is restricted to the gas flux in the protoplanetary disk. In sum, the total gas accretion rate can be expressed as 
\begin{equation}
    \dot{M}_{\rm p, g} = {\rm min} \left[ \left( \frac{{\rm d} M_{\rm p, g}}{{\rm d} t} \right)_{\rm KH}, \ \left( \frac{{\rm d} M_{\rm p, g}}{{\rm d} t} \right)_{\rm Hill}, \ \dot{M}_{\rm g} \right].
\end{equation}

\subsubsection{Planet migration}

Planets embedded in disks exchange angular momentum with the surrounding gas, resulting in their orbital migration, eccentricity and inclination damping.
We adopt a combined torque formula including both type I and type II regimes \citep{Kanagawa2018}:
\begin{equation}
    \Gamma  = f_{\rm tot} \Gamma_0 = \left[ f_{\rm I} f_{\rm s} + f_{\rm II} \left( 1-f_{\rm s} \right) \right] \Gamma_{\rm 0},
\end{equation}
where $\Gamma_{\rm 0} {=} M_{\rm p}^2 \, \Sigma_{\rm g} \, r^4 \, \Omega_{\rm K}^2/ M_{\star}^2 \, h_{\rm g}^2$ is the normalized torque strength, $f_{\rm I}$ and $f_{\rm II}$ are the type I and type II migration prefactors. The type II migration coefficient $f_{\rm II} {=} -1$ whereas the type I migration coefficient $f_{\rm I}$ is set by the disk thermal structure and local turbulent $\alpha_{\rm t}$ (see \cite{Paardekooper2011} for details). 
Differing from the traditional criterion within the viscous accretion disk framework \citep{Lin1986, Rafikov2002, Tanaka2002}, we employ $\alpha_{\rm t}$ instead of $\alpha_{\rm g}$ when addressing the gap opening and migration, motivated by the magnetohydrodynamic effect in the wind-driven disk.\citep{Aoyama2023}.
A smooth function of $f_s = 1/[1 + (M_{\rm p}/M_{\rm gap})^4]$ is chosen to avoid discontinuity and ensures that $\Gamma{\approx}\Gamma_{\rm I}$ when $M_{\rm p}{\ll} M_{\rm gap}$ and $\Gamma{\approx}\Gamma_{\rm I}/(M_{\rm p}/M_{\rm gap})^2$ when $M_{\rm p}{\gg} M_{\rm gap}$ \citep{Kanagawa2018}. 

We note that the heating torque from gas and pebble accretion \citep{BenitezLlambay2015, Masset2017, Cornejo2023} as well as other potential planet traps at the opacity transition regions such as ice-lines \citep{Kretke2012} are not taken into account in our study.

Planets in the inner viscously heated region can undergo outward migration ($f_{\rm I}>0$) when their masses are comparable to the optimal mass, which reads  
\begin{equation}
    {\rm M}_{\rm opt} = 0.9 \ \left(\frac{\alpha_{\rm t}}{10^{-3}}\right)^{2/3} \ \left(\frac{h_{\rm g}}{0.065}\right)^{7/3} \ \left(\frac{M_{\star}}{0.1 \ {\rm M}_{\odot}}\right) \ {\rm M}_{\oplus}.
    \label{eq:opt}
\end{equation}
A notable feature is that even though the protoplanets are distributed widely over the whole disk region, they would migrate convergently towards $r_{\rm tran}$ when reaching such a mass.

The inner disk is truncated by the stellar magnetospheric torque \citep{Lin1996, Liu2017} and the corresponding cavity radius is around $0.03$ au around young T Tauri stars around solar-mass. We set the inner disk boundary as $r_{\rm in}{=}0.01$ au around late dwarfs which is assumed to equal to their stellar-corotation radius with spin orbits of ${\approx}3$ days.  Any planets migrating interior to this radius are immediately stopped. In numerical integrations, we remove these planets inside the cavity radius to save the computational cost.
%and planets terminate their migration when entering the inner disk cavity. 

\subsection{Numerical setup}

We used numerical N-body simulations to study the growth and evolution of multi-protoplanets. We have employed the MERCURY code \citep{Chambers1999} with the Bulirsch-Stoer integrator. In the code planet-planet collisions are treated as inelastic mergers with conserved angular momentum when the separation of two planets is smaller than the sum of their physical radii. We ignore the influence of the potential energy released during the giant impact on the cooling and gas accretion. The mass of the remnant planet is a sum of both impactor and target.
Fragmentation and restitution \citep{Leinhardt2012, Mustill2018} are not considered in this work (see further discussions in Sect. \ref{sec:multi}). A planet with a distance greater than $100$ au from the central star is considered to be ejected from the planetary system.

The planet-disk interactions are implemented as accelerations:
\begin{equation}
    \boldsymbol{a_m} = -\dfrac{\boldsymbol{v}}{t_m}, \  \boldsymbol{a_e} = -2\dfrac{(\boldsymbol{v \cdot r})\boldsymbol{r}}{r^2 t_e}, \ \boldsymbol{a_i} = -\dfrac{\boldsymbol{v_z}}{t_i},
\label{eqn:a_mig}
\end{equation}
where $\boldsymbol{v}$ is the velocity vector. Modifying from \cite{Cresswell2008}, the migration, eccentricity, and inclination damping timescales are given by 
\begin{equation}
    t_{\rm m} =  \frac{t_{\rm wave}}{2 \, |f_{\rm tot}| \, h_{\rm g}^2}, \  t_{\rm e} = \frac{t_{\rm wave}}{0.78 \ |f_{\rm tot}|},  \  t_{\rm i} = \frac{t_{\rm wave}}{0.544 \ |f_{\rm tot}|},
\end{equation}
where 
\begin{equation}
   t_{\rm wave} = \frac{M_{\star}}{M_{\rm pl}} \ \frac{M_{\star}}{\Sigma_{\rm p} r^2} \ h_{\rm g}^4 \ \Omega^{-1}_{\rm K}.
   \label{eq:tim}
\end{equation}

In short,  the modified version of the code can handle planet-planet interactions and collisions and additionally account for the effects of  planet mass growth by pebble accretion, planet-gas disk interaction torques, and the corresponding eccentricities/inclinations damping.

\section{Growth of a single protoplanet}
\label{sec:single}

In this section we explore the growth and migration of a single protoplanet around a star of $M_{\star}{=} 0.1 \ M_{\odot}$. In our fiducial run  we assume the initial mass of the protoplanet to be $0.01 \ M_{\oplus}$ and the local turbulent diffusivity coefficient $\alpha_{\rm t}{=}10^{-3}$. The initial disk accretion rate is chosen as $\dot{M}_{\rm g,0} {=} 10^{-8} \ M_{\odot} \ \rm yr^{-1}$, and it starts to dissipate at $t_0{=}1$ Myr with a dispersal timescale $\tau_{\rm dep}$ of $0.5$ Myr. This corresponds to a disk lifetime $t_{\rm disk} {=}3.7$ Myr. The initial pebble flux is $\dot{M}_{\rm peb,0}{=} 3.3 \times  10^{-5} \ M_{\oplus} \rm yr^{-1}$, equivalently to $\xi_{p/g}{=} \dot{M}_{\rm peb}/\dot{M}_{\rm g}{=}1\% $.

We also investigate the influence of $\alpha_{\rm t}$ in Sect. \ref{subsec:viscosity}, pebble-to-gas mass flux ratio in Sect. \ref{subsec:metallicity}, disk lifetime in Sect. \ref{subsec:lifetime} and initial mass of protoplanet in Sect. \ref{subsec:mass}, respectively. 
The setup of disk and protoplanet parameters are listed in Table \ref{tab:param}.

\begin{table*}[h]
\caption{Disk and planet parameter setup in Sect. \ref{sec:single}.}
    \centering
    \begin{tabular}{c|c|c|c|c}
        \hline
         runs & $\alpha_{\rm t}$ & $\xi{=}\dot{M}_{\rm peb}/\dot{M}_{\rm g}$ & $t_{\rm disk}$ & $M_{\rm p0}$ \\ 
         & &  & (Myr) & ($M_{\oplus}$) \\
           \hline
     fiducial & $10^{-3}$ & $1\%$ & $3.7$ & $0.01$ \\ 
        disk turbulence & $10^{-4}$, $10^{-3}$, $10^{-2}$ & $1\%$& $3.7$ & 0.01\\
        disk solid mass & $10^{-4}$, $10^{-3}$, $10^{-2}$ & $2\%$ & $3.7$ & 0.01\\
       disk lifetime & $10^{-4}$, $10^{-3}$, $10^{-2}$ & $1\%$, $2\%$  & $6.3$ & $0.01$ \\
        protoplanet mass & $10^{-4}$, $10^{-3}$, $10^{-2}$ & $1\%$, $2\%$ & $3.7$ &  Eq. \ref{eq:Mp0}\\
%        & & & &\\
        \hline
    \end{tabular}
    \tablefoot{$t_{\rm disk}$ is calculated for the timespan when the gas surface density at 1 au drops to $1 \ {\rm g} \, {\rm cm}^{-2}$.}
    %\begin{tablenotes}
     %   \item Note: $t_{\rm disk}$ is calculated for the timespan when the gas surface density at 1 au drops to $1 \ {\rm g} \, {\rm cm}^{-2}$.
    %\end{tablenotes}
    \label{tab:param}
\end{table*}

\subsection{Fiducial case} \label{subsec:single}

\begin{figure}
    \centering
    \includegraphics[width=\columnwidth]{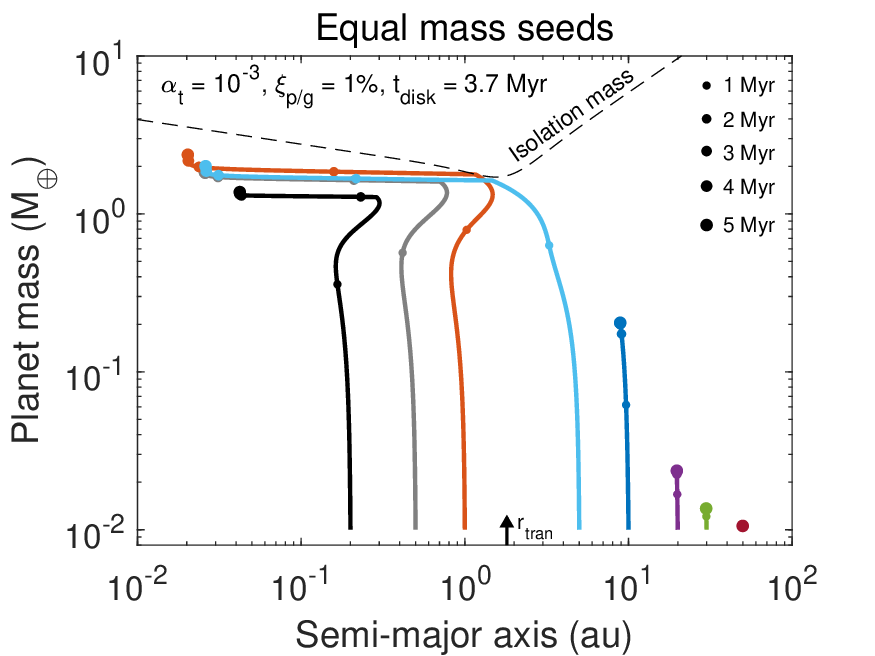}
    \caption{Growth and migration of individual protoplanets initiated at different disk locations around stars of $M_{\star}{=}0.1 \ {M}_{\odot}$. The dashed line represents the pebble isolation mass at $t{=}1.3$ Myr. This is when the fastest growing planet reaches its isolation mass. The arrow indicates the disk transition radius, and the increasing sizes of the dots denote the disk evolution at one Myr intervals. The planet attains the highest mass at a moderate birth radial distance close to the transition radius. The planet and disk parameters are referred to Table \ref{tab:param}.}
    \label{fig:eqsingle1}
\end{figure}

Figure \ref{fig:eqsingle1} illustrates the growth of individual protoplanets at various birth locations $r_0$. The dashed line refers to $M_{\rm iso}$ at the time when the fastest growing protoplanet reaches ($t{=}1.3$ Myr), and the vertical arrow indicates the transition radius between two disk heating sources. The increasing size of the dots represents the time evolution, with intervals of a Myr.   

The red curve in Figure \ref{fig:eqsingle1} depicts the growth of a protoplanet at $r_0{=}1$ au. The mass of the protoplanet increases by two orders of magnitudes through pebble accretion during the first Myr.  As the planet reaches $M_{\rm opt}{\sim} 0.5 M_{\oplus}$, it starts to migrate outward to $r_{\rm tran}$ due to a strong, positive corotation torque in the viscously heated disk region \citep{Liu2019a}. However, this corotation torque gradually diminishes as disk dissipating and the planet mass further increasing, causing rapid inward migration. The planet reaches $M_{\rm iso}{=}1.8 \ M_{\oplus}$ at $t{=}1.3$ Myr and $r{=}1.2$ au. Because of a relatively low core mass, it since then only accretes a limited amount of gas and eventually grows into a close-in, super-Earth planet of $M_{\rm p}{=} 2.4 \ M_{\oplus}$. 

The growth differs when the protoplanets are born at different $r_0$. Only protoplanets with $r_0{\sim}r_{\rm tran}$ can grow sufficiently massive and undergo large-scale radial migration. We term this region that protoplanets can grow beyond $0.5 \ M_{\oplus}$ as the efficient planet growth region. Protoplanets with initial closer-in and further-out orbits end up as lower-mass planets. This $r$-dependent mass growth correlates with the gas disk scale height, which governs the efficiency of pebble accretion. In the $2$D accretion case, a larger gas scale height indicates a faster headwind speed. The pebbles drift too fast and are less likely to be accreted by the planets. On the other hand, in the $3$D accretion case, a larger gas scale height also means more vertically extended pebble layers, leading them less efficient to be attracted by the planet. Taken together, the highest efficient pebble accretion occurs when the disk scale height has the lowest value. This corresponds to the disk location at $r_{\rm tran}$, since $h_{\rm vis} {\propto} r^{-1/16}$ in the inner disk and $h_{\rm irr} {\propto} r^{2/7}$ in the outer disk (see Eqs. \ref{eq:hvis} and \ref{eq:hirr}). As a result, the planets exhibit a peak growth rate at $r{\sim}r_{\rm tran}$. However, none of these protoplanets finally grow into massive, gas-dominated planets, due to the fact that their core masses are too low to initiate rapid gas accretion.

\subsection{Disk turbulence} \label{subsec:viscosity}

It is expected that disks are turbulent, which dynamically stirs up solid particles and affects the pebble accretion efficiency of planets \citep{Johansen2017}, as well as the planetesimal formation through the streaming instability \citep{Johansen2014, Drazkowska2023}. One direct method to assess the strength of the turbulence is by deriving the turbulence-induced broadening observed in molecular line emissions \citep{Najita1996}. Instead of a universal turbulent viscosity, the value of $\alpha_{\rm t}$ varies from one disk to another \citep{Flaherty2018, Flaherty2020, Teague2018}. The inner disk region and upper layer of the source SVS 13 shows supersonic turbulence \citep{Carr2004}, while HD 163296 demonstrates moderate levels of turbulence with $\alpha_{\rm t} < 2.5 \times 10^{-3}$ \citep{Flaherty2015, Flaherty2017}.

Another approach to constrain turbulence is through geometric considerations \citep{Rosotti2023}, for example the dust vertical extent or the radial width of disks influenced by settling/radial drift and turbulence diffusion \citep{Whipple1972, Pinte2016, Rosotti2020}. Observations of Oph 163131 \citep{Villenave2022} and the DSHARP survey \citep{Andrews2018, Dullemond2018} suggest a preference for low turbulent viscosity of $\alpha_{\rm t} \lesssim 10^{-4}$ to moderate values of $\alpha_{\rm t} \sim 10^{-3}$. 
%Studies on disk populations also indicate a median $\alpha_{\rm t}$ ranging from $3 \times 10^{-4}$ to $3 \times 10^{-3}$\citep{Ansdell2018, Trapman2020}. 
Overall, turbulent viscosity typically ranges from $\alpha_{\rm t} = 10^{-2}$ to $10^{-4}$ in different disks, leading us to investigate the influence of disk turbulence on planet growth and migration within this parameter space.

\begin{figure*}
    \centering
    \includegraphics[width=\linewidth]{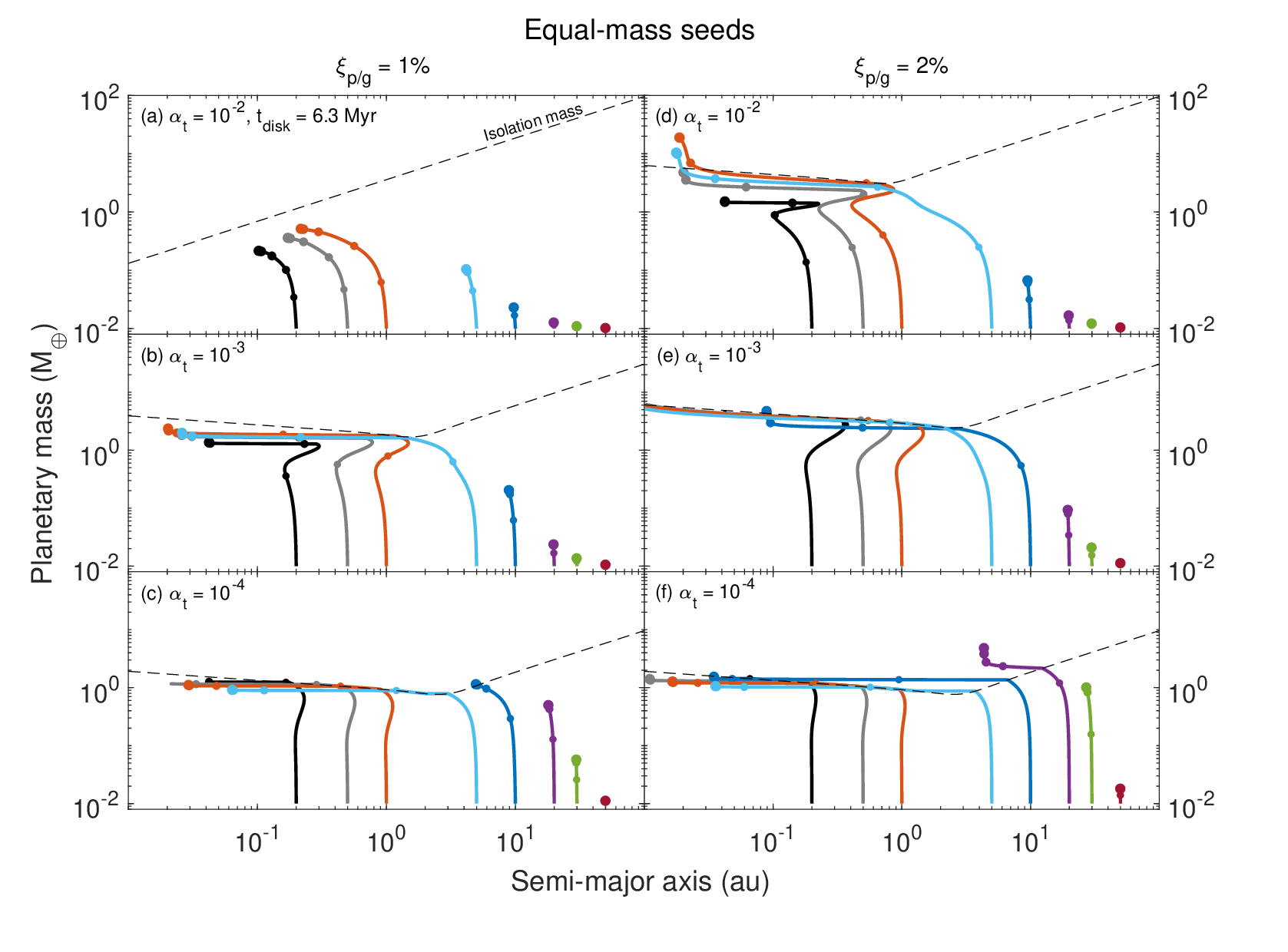}
    \caption{
    Growth and migration of single protoplanets born with lunar masses at different turbulent levels and pebble-to-gas mass flux ratios around stars of $M_{\star}{=}0.1 \ {M}_{\odot}$. Three turbulent coefficients of $\alpha_{\rm t}=10^{-2}$ (upper), $10^{-3}$ (middle) and $10^{-4}$ (lower) and two pebble-to-gas flux ratios of $\xi_{\rm p/g}{=} 1\%$ (left) and $2\%$ (right) are shown. The planet and disk parameters are listed in Table \ref{tab:param}.
  The dashed line represents the pebble isolation mass at the time when the planet born at $1$ au reaches this value. Note that in panel (a) planets never approach the isolation mass. We instead adopt the isolation mass using the stellar irradiation model. The increasing sizes of the dots denote the disk evolution at one Myr intervals. Massive planets prefer to form in mental-rich and highly turbulent disks.
    }
    \label{fig:eqsingle}
\end{figure*}

We maintain a constant global disk angular momentum transport efficiency $\alpha_{\rm g}$, and the results are depicted in Figure \ref{fig:eqsingle}. In highly turbulent disks with $\alpha_{\rm t}{=}10^{-2}$, pebbles are vertically extended over the gas scale height, leading to a suppression of pebble accretion compared to the fiducial run. In addition, the pebble isolation mass increases with disk turbulence (Eq. \ref{eq:Miso}), making planets more challenging to reach $M_{\rm iso}$ before disk dissipation. The maximum mass that a planet can attain is approximately Venus-mass in Fig. \ref{fig:eqsingle}a. On the other hand, in weakly turbulent disks with $\alpha_{\rm t}{=}10^{-4}$, planet growth speeds up due to efficient pebble accretion. Planets can grow up to $M_{\rm iso}$ within 1 Myr at $r_{0}{\sim}r_{\rm tran}$. However, in such a case $M_{\rm opt}$ is lower, and the effect of outward migration is insignificant. Planets migrate rapidly into the inner disk region. Moreover, $M_{\rm iso}$ is also lower, and planets are prevented from accreting substantial gas to become gas giants. Only small planets with the highest mass of ${\sim} 1 \ M_{\oplus}$ form in the end.

In brief, the growth of massive planets from a single protoplanet is largely impeded,  in the disks with either very high or very low turbulent levels.

\subsection{Pebble-to-gas mass flux ratio} \label{subsec:metallicity}

The total solid mass in disks, quantified by $\xi_{p/g}$, the pebble-to-gas mass flux, crucially determines the planet growth timescale. A higher pebble mass flux facilitates the formation of massive planets.

The right panel of Figure \ref{fig:eqsingle} illustrates the growth of protoplanets in metal-rich disks with a pebble-to-gas flux ratio $\xi_{p/g}$ of $ 2\%$. The efficient planet growth region is wider, and protoplanets at further-out disk regions can grow more quickly and substantially in this situation compared to the case of the nominal $\xi_{p/g}{=}1\%$ (left panel of Figure \ref{fig:eqsingle}). For instance, super-Earth can form in disks with $\alpha_{\rm t} = 10^{-3}$ at $r_0{=} 10$ au and $\alpha_{\rm t}{=}10^{-4}$ disks at $r_0{=} 20$ au in Figure \ref{fig:eqsingle}e and f, respectively, because of their considerable core masses.

Notably, protoplanets in highly turbulent disks experience even more significant mass growth due to the fact that less efficient pebble accretion is largely compensated by a large supply of the pebble reservoir (Figure \ref{fig:eqsingle}d). In such a case, the optimal mass is higher, allowing them to retain at $r_{\rm tran}$ for a longer time to proceed the mass growth. The pebble isolation mass is also higher. Once they reach such a massive core, they are more likely to initiate rapid gas accretion. Evidently, in Figure \ref{fig:eqsingle}d protoplanets born at $r{<}5$ au reach $M_{\rm iso}{\sim}3 \ M_{\oplus}$ at a relatively early time and eventually grow into Neptune-mass planets. 

To conclude, in metal-rich disks, high disk turbulence may no longer pose a threat to the formation of massive planets. The true barrier is $M_{\rm iso}$, which determines the ability of runaway gas accretion.   

\subsection{Disk lifetime} \label{subsec:lifetime}

%Observations in young, nearby star-forming regions indicate that low-mass M dwarfs generally harbor a higher fraction of disks compared to their solar-type counterparts (Ribas et al. 2014, 2015; Sicilia-Aguilar et al. 2009).

\begin{figure*}
    \centering
    \includegraphics[width=\linewidth]{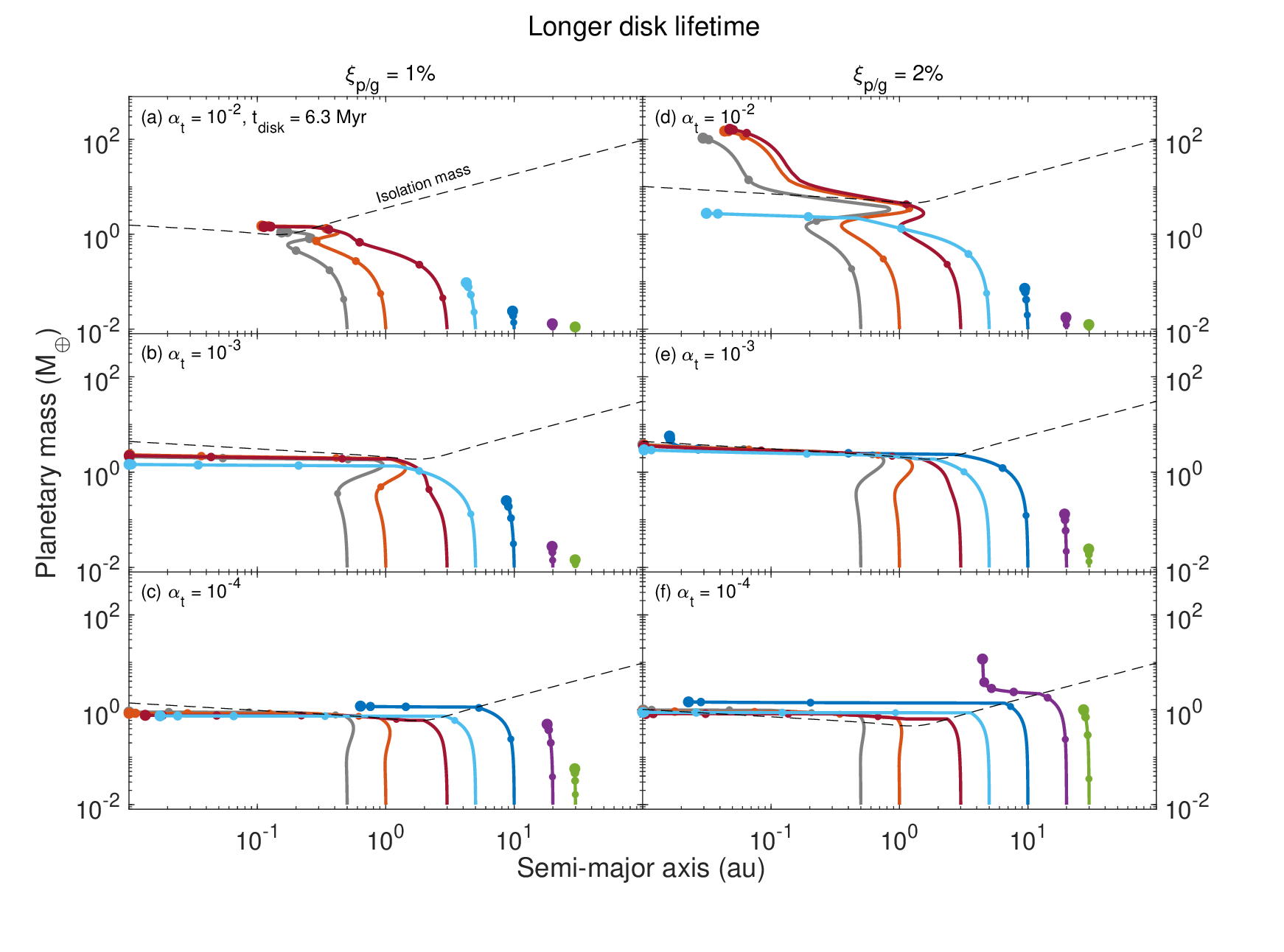}
    \caption{
        Growth and migration of single protoplanets in disks of a long lifetime  at different turbulent levels and pebble-to-gas mass flux ratios around stars of $M_{\star}{=}0.1 \ {M}_{\odot}$. Three turbulent coefficients of $\alpha_{\rm t}=10^{-2}$ (upper), $10^{-3}$ (middle) and $10^{-4}$ (lower) and two pebble-to-gas flux ratios of $\xi_{\rm p/g}{=} 1\%$ (left) and $2\%$ (right) are shown. The planet and disk parameters are listed in Table \ref{tab:param}.
  The dashed line represents the pebble isolation mass at the time when the planet born at $1$ au reaches this value. The increasing sizes of the dots denote the disk evolution at one Myr intervals. Compared to \Fg{eqsingle}, more massive planets from in disk with a longer lifetime.}
    \label{fig:disksingle}
\end{figure*}

Giant planets assemble gas and solids within a finite protoplanetary disk lifetime. Therefore, the survival time of the gaseous disk is expected to play a decisive role. Here we keep the total gas disk mass the same as the fiducial run and investigate the influence of a longer disk lifetime on planet growth and migration. This long-lived disk is characterized by a lower initial disk accretion rate $\dot{M}_{\rm g, 0} {=} 6 \times 10^{-9} \ M_{\odot} \ {\rm yr}^{-1}$ and an extended disk dissipation such that $t_0 {=} 1.5 \ \rm Myr$ and $\tau_{\rm dep} {=} 1.0 \ \rm Myr$. The disk lifetime is $6.3$ Myr in this case. Owing to the slow disk dissipation $\dot{M}_{\rm g}$ becomes higher than the fiducial case after $t{=}1.3$ Myr. The results are presented in Figure \ref{fig:disksingle}.

 Since the pebble flux is attached to the gas flux, protoplanets grow slowly in the early stage, and they approach $M_{\rm iso}$ at later times (generally later than $2$ Myr) in Figure \ref{fig:disksingle} compared to Figure \ref{fig:eqsingle}. The disk mass also dissipates slower. After $t{=}1.3$ Myr, both $M_{\rm iso}$ and $\dot M_{\rm peb}$ are higher in long-lived disks. As such, protoplanets speed up their growth at these advanced phases and reach a higher core mass. Meanwhile, $M_{\rm opt}$ is also higher in disks with a relatively high $\dot{M}_{\rm g}$. Outward migration is also more profound in Figure \ref{fig:disksingle}d when the planet grows beyond a few Earth masses.

Importantly giant planet formation is significantly promoted in long-lived, highly turbulent, and metal-rich disks, as shown in Figure \ref{fig:disksingle}d.

\subsection{Protoplanets formed by streaming instability} \label{subsec:mass}

\begin{figure*}
    \centering
    \includegraphics[width=\linewidth]{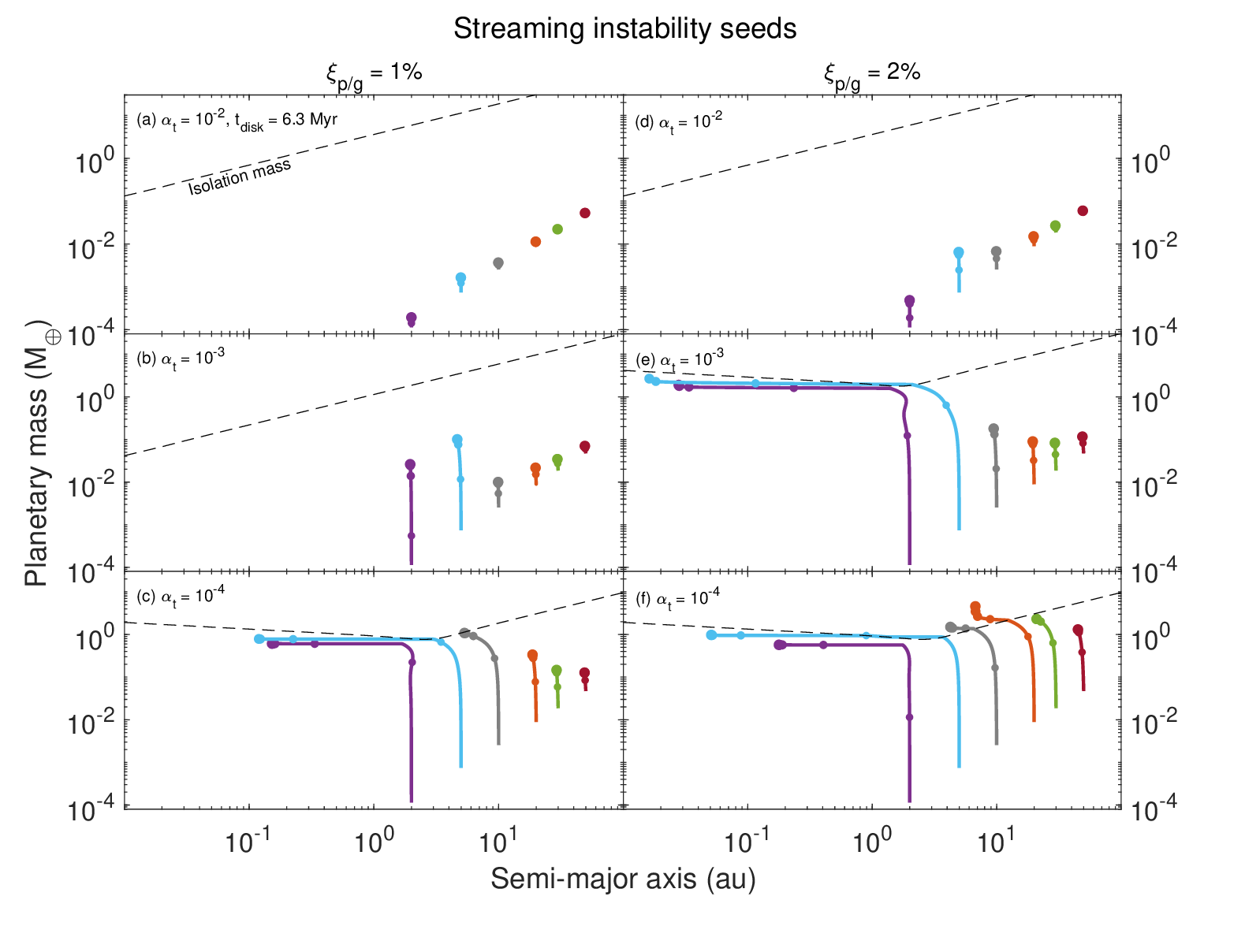}
    \caption{ Growth and migration of single protoplanets formed by streaming instability at different turbulent levels and pebble-to-gas mass flux ratios around stars of $M_{\star}{=}0.1 \ {M}_{\odot}$. Three turbulent coefficients of $\alpha_{\rm t}=10^{-2}$ (upper), $10^{-3}$ (middle) and $10^{-4}$ (lower) and two pebble-to-gas flux ratios of $\xi_{\rm p/g}{=} 1\%$ (left) and $2\%$ (right) are shown. The planet and disk parameters are listed in Table \ref{tab:param}. Compared to \Fg{eqsingle}, the growth of the planets is significantly impeded unless in disks with low turbulence and high pebble flux.}
    \label{fig:SIsingle}
\end{figure*}

We assume that the protoplanets form by the streaming instability mechanism. Unlike previous circumstances where protoplanets had an equal lunar mass, the mass derived from Eq. \ref{eq:Mp0} correlates with $M_{\star}$, $\Sigma_{\rm g}$, and $h_{\rm g}$, increasing with $r$ within a range of $10^{-4} \ M_{\oplus}$ to approximately $0.1 \ M_{\oplus}$ \citep{Liu2020b}. We note that the streaming instability triggering condition also correlates with disk properties such as the local disk metallicity \citep{Yang2017, Li2021}. We assume that even though the global disk metallicity is below the threshold value, the local solid density can still be enhanced to fulfill the streaming instability by various of hydrodynamical and magnetic instabilities (see references in \cite{Lenz2019}). Following this, the mass of forming planetesimal is degenerate from the global $\xi_{\rm p/g}$.   We perform simulations with disk parameters identical to the fiducial run, and the results are demonstrated in Fig. \ref{fig:SIsingle}.

Compared to Figure \ref{fig:eqsingle}, protoplanets formed by the streaming instability face significant challenges in growing masses in highly turbulent disks. This situation holds both true for disks with $\xi_{\rm p/g}{=}1\%$ and $2\%$ (Fig. \ref{fig:SIsingle}a and d). In a disk with $\alpha_{\rm t} {=} 10^{-3}$, the region of efficient planet growth spans approximately $2{-}5$ au (Fig. \ref{fig:SIsingle}e), narrower than that in equal-mass cases. The masses of the planets increase by two orders of magnitude at $\xi_{\rm p/g}{=}1\%$ (Fig. \ref{fig:SIsingle}b), whereas the formation of Earth-sized planets becomes feasible at $\xi_{\rm p/g}{=}2\%$ (Fig. \ref{fig:SIsingle}e). The outcome is natural to understand from the difference in initial protoplanet masses. The mass from streaming instability is generally lower than lunar mass within $20$ au. Therefore, protoplanets with lower masses have lower gravitational potential to capture pebbles, resulting in a longer growth time. 

Massive planets are more likely to grow in disks with low $\alpha_{\rm t}$ and high $\xi_{\rm p/g}$. We find that the general growth pattern is similar, but protoplanets located at the outer disk region in Fig. \ref{fig:SIsingle}f can attain slightly higher masses than those in Fig. \ref{fig:eqsingle}f since $M_{\rm p0}$ there are higher than $0.01 \ M_{\oplus}$. But the growth is still limited and only cold, super-Earth planets form eventually.  

In conclusion, the formation of massive planets is more difficult when protoplanets are assumed to form by the streaming instability rather than being born with an equal lunar mass.

\section{Growth of multi-protoplanets}
\label{sec:multi-protoplanets}
In the previous section we present the growth of a single protoplanet under various disk and planet parameters. However, multi-planetary systems are commonly observed. It is essential to understand how the formation and evolution of the planetary system from more realistic configurations started from multi-protoplanets.

In order to investigate this, we conduct N-body numerical simulations that account for the gravitational interactions among multiple protoplanets. We test whether the growth pattern differs in single and multi-protoplanet cases. The illustrations of the N-body simulations are presented in Section \ref{sec:multi}, and we discuss their outcomes in a parameterized manner in Section \ref{sec:para_space}.

For the multi-protoplanet cases, we start with $N{=}20$ protoplanets initially. Since our goal is to explore the possibility of giant planet formation, we place these protoplanets within the efficient planet growth zone that we explored in our single-planet growth study (Section \ref{sec:single}). Note that the radial width of the zone varies among different parameter setups. We randomly select the separation between protoplanets from $10$ to $50$ mutual Hill radius to fill all bodies within this zone. We test that the final outcome is not sensitive to the choice of their mutual separations, as long as they are well separated at the beginning. For comparison, we also plot the single protoplanet growth by optimizing $r_0$ to let the planet reach the highest mass.   

We also explore a few cases where $N{=}30$. However, due to the gravitational interactions and orbital excitations, there is always a limited number of planets that can grow sufficiently massive and dominate the subsequent dynamical evolution. As a result, the final masses and numbers of giant planets are not strongly dependent on the adoption of $N$ from $20$ to $30$ \citep{Emsenhuber2021}. However, increasing $N$ would significantly increase the computational time. We thus limit our multi-protoplanet explorations to $N{=}20$.

The initial eccentricities and inclinations of the protoplanets follow the Rayleigh distributions, with a scaled eccentricity and inclination of $e_0{=}2i_0{=}0.01$. We also randomize the initial phase angles of these bodies. The simulations are terminated when the disks are fully dissipated ($5$ Myr for the fiducial disks and $10$ Myr for the disks with a longer lifetime).

\subsection{Illustration runs} \label{sec:multi}
\Fg{eqmulti} is an example that demonstrates the growth and migration of multiple protoplanets. The planets that survived after $5$ Myr are depicted with colored lines, while those ejected or merged are represented by grey lines.

\begin{figure*}
    \centering
    \includegraphics[width=\linewidth]{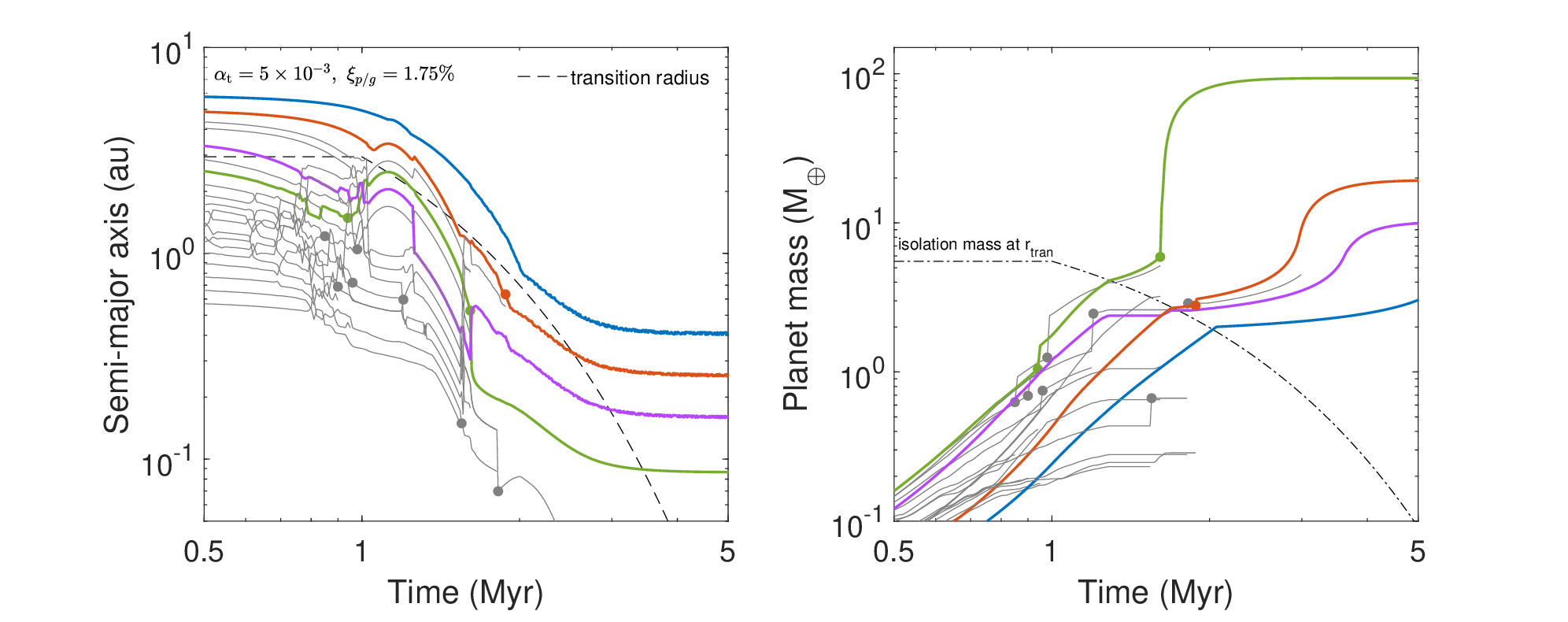}
    \caption{Semi-major axis (left) and mass (right) evolution from multi-protoplanet growth. The planet and disk parameters are: $M_{\rm p0} =0.01 \ M_{\oplus}$, $\alpha_{\rm t}{=}5 \times 10^{-3}$, $M_{\rm d}{=}0.15 \ M_{\star}$, $\xi_{\rm p/g}{=}1.75\%$ and $t_{\rm disk}{=}3.7$ Myr. The filled dots indicate the planet-planet collisions. The transition radius is illustrated in the dashed line in the left panel, while the dot–dashed lines in the right panel represent the pebble isolation mass at the transition radius. By a combination of pebble accretion and planet-planet collisions, a system with one gas giant and three super-Earths form in very low-mass stars of $M_{\star}{=}0.1 \ M_{\odot}$. }
    \label{fig:eqmulti}
\end{figure*}

Initially, protoplanets are widely separated and their masses increase through pebble accretion. Due to the heterogeneity in growth rates at different $r_0$, protoplanets near $r_{\rm tran}$ (dashed line in the left panel of Fig. \ref{fig:eqmulti}) acquire higher masses than others. Once their masses exceed ${\sim}0.1 \ M_{\oplus}$, they undergo inward migration. The continued mass increase and convergent migration lead to the compression of these bodies' orbits, triggering dynamical instabilities and frequent planet-planet collisions (denoted by dots in Fig. \ref{fig:eqmulti}) at $t{\sim}1$ Myr. 

The outcome of a close encounter between two planets can be determined by the ratio of the surface escape velocity $v_{\rm esc}$ and the escape velocity of the planetary system (${=}\sqrt{2}v_{\rm K}$ where $v_{\rm K}$ is the Keplerian velocity at the planet location). This can be calculated by \citep{Goldreich2004}
\begin{equation}
    \begin{aligned}
  \Lambda^2  & = \frac{v_{\rm esc}^2}{v_{\rm K}^2} = \left(\frac{M_{\rm p}}{M_{\star}} \right) \left(\frac{r}{R_{\rm p}} \right)\\
  & {\simeq} 0.25 \left( \frac{r} {1 \ {\rm au}}\right) \left( \frac{\rho_{\rm planet}} {2 \ {\rm g \ cm^{-3}}} \right) \left( \frac{R_{\rm p}} {1 \ R_{\oplus}}\right)^{2}  \left( \frac{M_{\star}} {0.1 \ M_{\odot}}\right)^{-1}.
    \end{aligned}
\end{equation}
The planet bulk density $\rho_{\rm planet}{\sim}2 \ \rm g \ cm^{-3}$ since they form beyond the water-ice line. We also find that more massive planets collide at closer-in orbits with generally $\Lambda{<}1$. As such, collisions between close-encounters are favored rather than ejections.  
At this stage the planets of $M_{\rm p}{\sim}M_{\oplus}$ at $r {\sim} 1$ au have moderate eccentricities of $0.1$. So the impact velocity among these planets approximates ${\sim} e v_{\rm K}{\sim} 1 \ {\rm km/s}$, lower than their escape velocity.  In this regime the accretion efficiency is very high (see Fig. 6 of \cite{Cambioni2019}) and the perfect merger treatment is therefore appropriate \citep{Asphaug2010}.   
The collisional timescale is given by $\tau_{\rm col}{\sim} (n  \sigma_{\rm col} \Delta v )^{-1}$, where $n= N/(2 \pi r \Delta r \Delta z)$ is the planet number density, $r{\sim}3$ au and $\Delta r {\sim}3$ au and $\Delta z {\sim} i  \times  r$, $\sigma_{\rm col}{\sim} \pi R_{\rm p}(1 + v_{\rm esc}^2/\Delta v^2) $ is the collisional cross section, and $\Delta v{\sim} e v_{\rm K}$ is the relative velocities among planets. We can  estimate $\tau_{\rm col}{\sim} 0.3$ Myr for the planets with $R_{\rm p}{=}1 \ R_{\oplus}$ and $e{\sim}2i{\sim}0.1$, in agreement with the results shown in Fig. \ref{fig:eqmulti}.

After a series of mergers, a few protoplanets double their masses, which boosts their subsequent pebble accretion. These bodies, with masses around $M_{\rm p}{\sim} M_{\rm opt}$, begin outward migration, leading to chaotic orbits of planets close to $r_{\rm tran}$. This triggers a second phase of strong perturbations and planet-planet collisions at $t{\sim}1.5{-}2$ Myr. Collisions in this phase are not as intense as the first one, since the number of planets in the system has been reduced. 

As the gas disk gradually dissipates, the pebble isolation mass drops below ${\sim} 3 \ {\rm M}_{\oplus}$ at $t{\sim}1.5$ Myr close to the transition radius. Only a few massive bodies remain in the system after multiple planet-planet collisions and scatterings. 
The largest body reaches the core mass of ${\sim}11 \ M_{\oplus}$ after a collision at $\sim 1.6$ Myr. It initaites runaway gas accretion and quick becomes a giant planet with a mass of $M_{\rm p}{=}100 \ M_{\oplus}$ and an orbital period of 30 days.   
The other three lower mass bodies attain $M_{\rm iso}$ at later times, acquiring  the residual disk gas and  growing into super-Earth planets. 
All these planets undergo inward migration, and end up in final orbits of $r{\sim}0.1{-}0.4$ au. The outermost three planets are trapped into $4$:$2$:$1$ mean motion resonances.

% Need one paragraph to compare with singles

\subsection{Parameter survey} \label{sec:para_space}

In order to validate our hypothesis that the presence of multiple planet-planet collisions promotes giant planet formation, we conduct an extensive parameter study using N-body simulations by varying two key disk parameters: turbulent level and total solid mass. The solid disk mass correlates with $\xi_{\rm p/g}$, which is calculated by integrating the pebble mass flux over the disk's lifetime.

We employ a $5\times 5$ grids to explore the ranges of $\alpha_{\rm t}$ and $\xi_{\rm p/g}$. The simulations are conducted at the boundaries where $\alpha_{\rm t} = 10^{-4}, 5 \times 10^{-4}, 10^{-3}, 5 \times 10^{-3}, 10^{-2}$ and solid disk mass of $50$, $63$, $75$, $88$ and $100 \ M_{\oplus}$. The intermediate region is populated using linear interpolation.  %$\xi_{\rm p/g} = 1\%, 1.25\%, 1.5\%, 1.75\%$ and $2\%$, 
In order to account for the statistical nature of multi-planet interactions, we perform five sets of N-body simulations by randomizing their initial mutual separations and orbital phase angles at each point.
The final planet mass is adopted from the largest forming planets over these five realizations.

We discuss the results and implications of the runs with fiducial parameters in Sect. \ref{sec:psfiducial},  longer disk lifetime in Sect. \ref{sec:pstimescale} and initial mass of protoplanet from streaming instability in Sect. \ref{sec:psmass}, respectively. The planet and disk parameters are provided in table \ref{tab:param2}.

\begin{table*}[h]
\caption{Disk and planet parameters in Section \ref{sec:multi-protoplanets} and Section \ref{sec:discussion} }
    \centering
    \begin{tabular}{l|c|c|c|c|c|c|c|c}
        \hline
         runs & $\alpha_{\rm t}$ & $\xi_{\rm p/g}{=}\dot{M}_{\rm peb}/\dot{M}_{\rm g}$ & $t_{\rm disk}$ & $\tau_{\rm s}$ & $M_{\rm p0}$ & $M_{\star}$ & $L_{\star}$ & disk \\ 
         & &  & (Myr) & & ($M_{\oplus}$) & ($M_{\odot}$) &  ($L_{\odot}$) & structure\\
           \hline
     fiducial (Sect. \ref{sec:multi}  and  \ref{sec:para_space})  & $10^{-4}{-}10^{-2}$  & $1{-}2\%$ & $3.7$ & 0.05 & $0.01$ & $0.1$ & $0.01$ & vis + irr\\ 
       long disk lifetime (Sect. \ref{sec:para_space}) & $10^{-4}{-}10^{-2}$ & $1{-}2\%$ & $6.3$ & 0.05 & $0.01$ & $0.1$ & $0.01$ & vis + irr\\
        SI protoplanet mass (Sect. \ref{sec:para_space}) & $10^{-4}{-}10^{-2}$ & $1{-}2\%$ & $3.7$ & 0.05 &  Eq.\ref{eq:Mp0} & $0.1$ & $0.01$ & vis + irr \\
        $\tau_{\rm s}$ limited by fragmentation (Sect. \ref{sec:stokes}) & $10^{-4}{-}10^{-2}$  & $1{-}2\%$ & $3.7$ & $0.05 \times (10^{-3}/\alpha_{\rm t}) $ & $0.01$ & $0.1$ & $0.01$ & vis + irr \\
 only stellar irradiation disk  (Sect. \ref{sec:disk})  & $10^{-4}$ & $2\%$ & $3.7$ & 0.05 &  $0.01$ & $0.1$ & $0.01$ &  irr \\
 high stellar luminosity (Sect.  \ref{sec:luminosity}) & $10^{-4}$ & $2\%$ & $3.7$ & 0.05 &  $0.01$ & $0.1$ & $0.1$ & vis + irr \\
   high stellar mass (Sect. \ref{sec:starmass}) & $10^{-4}{-}10^{-2}$ & $1{-}2\%$ & $3.7$ & 0.05 & $0.01$ & $0.2$ & $0.04$ & vis + irr \\
%        & & & &\\
        
        \hline
    \end{tabular}
    \tablefoot{We note that $\xi_{\rm p/g}$ of $1{-}2\%$ corresponds to pebble disk mass of $50{-}100 \ M_{\oplus}$ around stars of $M_{\star}{=}0.1 \ M_{\odot}$ and $100{-}200 \ M_{\oplus}$ around stars of $M_{\star}{=}0.2 \ M_{\odot}$.}
    \label{tab:param2}
\end{table*}

\subsubsection{Fiducial case} \label{sec:psfiducial}

\begin{figure*}
    \centering
    \includegraphics[width=\linewidth]{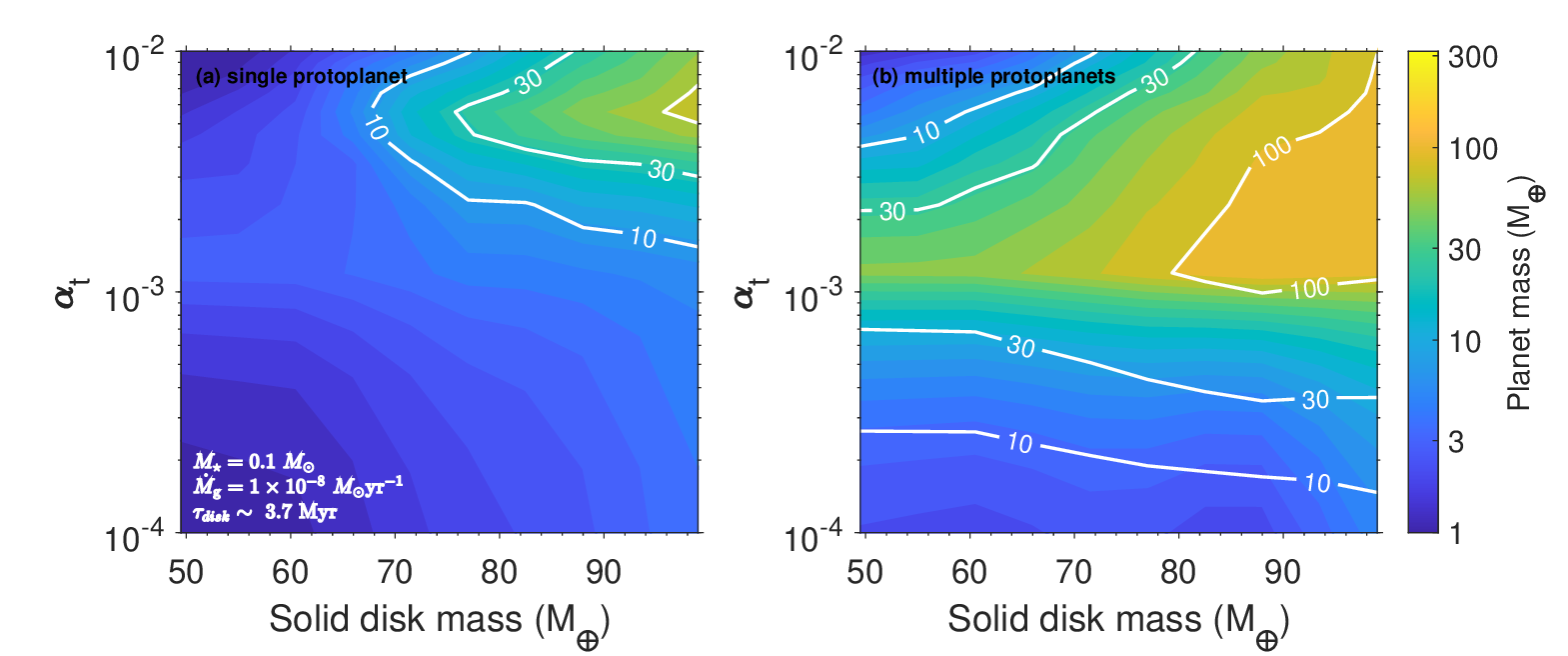}
    \caption{Formation of massive planets in the equal protoplanet mass scenario as a function of disk turbulent level and solid disk mass. The left and right panels illustrate the single protoplanet and multi-protoplanet cases. The protoplanets are assumed to be equal lunar mass and the disk lifetime is $3.7$ Myr. Other model parameters are listed in  Table \ref{tab:param}.  The colorbar gives the resulting planet with the highest mass, while the contour lines indicate masses of $10$, $30$, and $100$, respectively. Compared to the single protoplanet case, the parameter ranges for the formation of giant planets are wider in the multi-protoplanet case.  }
    \label{fig:eqparam}
\end{figure*}

Fig. \ref{fig:eqparam} displays the highest masses that planets can attain through the growth and evolution of either a single protoplanet (left) or multiple protoplanets (right). The color denotes the final planet mass, with yellow representing the most massive planets and blue representing the lightest ones. The white lines  correspond to planet masses of $10$, $30$, and $100 \ M_{\oplus}$, respectively.

In the simulations of single protoplanets, gas giant planets exclusively form in disks with $\alpha_{\rm t} {\sim} 0.5\times 10^{-3}{-} 10^{-2}$ and total solid mass of $100 \ M_{\oplus}$ ($\xi_{\rm p/g}{\sim}2\%$). This is because $M_{\rm opt}$ is higher in the moderate and high-turbulent disks, allowing the planet to retain a long time outside before rapid inward migration. 
Giant planets with orbital period up to 100 days form in our model (an illustration simulation is shown in Fig. \ref{fig:eqmulti2}).
Besides, a higher $M_{\rm iso}$ in these higher $\alpha_{\rm t}$ cases means that planets have the ability to reach more massive solid cores, facilitating subsequent gas accumulation.

Yet, the drawback is that pebble accretion efficiency is lower in such high-turbulent disks. Therefore, giant planet formation only succeeds when disks have a massive supplier of pebble reservoir (high $\xi_{\rm p/g}$). This is why massive planets only occur in the yellow region on the right corner of Fig. \ref{fig:eqparam}a. Alternatively, in cases of low turbulent disks, planets are incapable of growing massive, as $M_{\rm iso}$ is too low to initiate rapid gas accretion.

In the simulations starting with multi-protoplanets, we find that the giant planet formation zone becomes wider in $\alpha_{\rm t}{-}\xi_{\rm p/g}$ space and shifts towards lower turbulence and less massive disks (yellow region in Fig. \ref{fig:eqparam}b). For disks with $\alpha_{\rm t}{=} 10^{-3}$, $M_{\rm iso}$ at $r{=}1$ au is approximately $3 \ {\rm M}_{\oplus}$. However, planets with a core mass of $M_{\rm iso}$ fail to grow massive within the disk lifetime (see Fig. \ref{fig:eqsingle}). Nevertheless, their core mass can be further increased by planet-planet collisions. After a few giant impacts, planets with a core mass of $6{-}8 \ M_{\rm iso}$ can rapidly accrete gas, leading to giant planet formation (see \fg{eqmulti}). We find that the growth of giant planets of $M_{\rm p} {>} 50 \ M_{\oplus}$ is feasible when disks have more than $70 \ M_{\oplus}$ pebbles at $\alpha_{\rm t}{\gtrsim} 10^{-3}$ (Fig. \ref{fig:eqparam}b). 

To conclude, compared to the growth of a single protoplanet, giant planet formation is more pronounced in the presence of multiple protoplanets by considering their subsequent convergent migration and planet-planet collisions.

\subsubsection{Disk lifetime} \label{sec:pstimescale}

\begin{figure*}
    \centering
    \includegraphics[width=\linewidth]{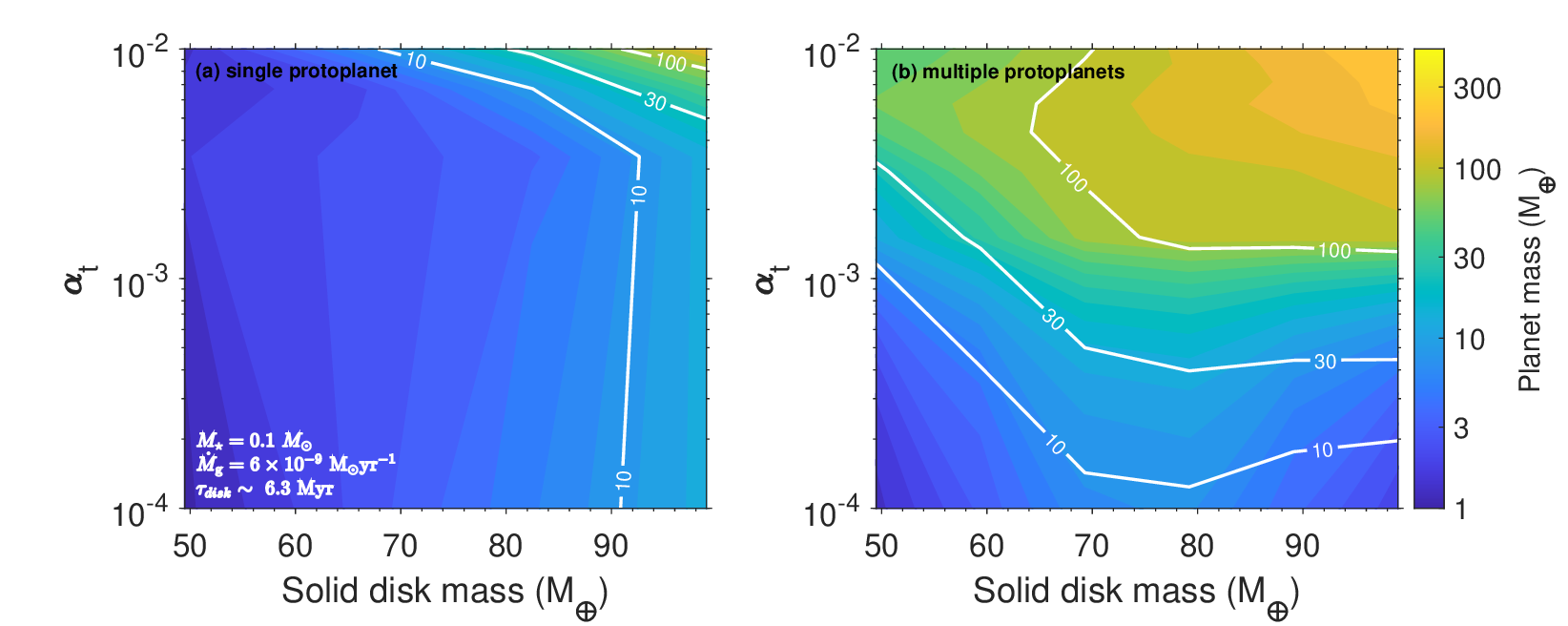}
    \caption{Formation of massive planets in the long disk lifetime scenario as a function of disk turbulent level and solid disk mass. The left and right panels illustrate the single protoplanet and multi-protoplanet cases.  The protoplanets are assumed to be equal lunar mass and the disk lifetime is $6.3$ Myr. Other model parameters are listed in  Table \ref{tab:param}.  The colorbar gives the resulting planet with the highest mass, while the contour lines indicate masses of $10$, $30$, and $100$, respectively. Giant planets are more likely to occur in disks with longer lifetimes.  }
    \label{fig:Lparam}
\end{figure*}

We explore the impact of longer disk lifetime on planet growth and the parameter map is given in  Figure \ref{fig:Lparam}. 
Compared to previous runs with a short disk lifetime, we find in Fig.  \ref{fig:Lparam}a that the giant plant formation zone becomes narrower and only peaks at higher $\alpha_{\rm t}$ and $\xi_{\rm p/g}$ in the single protoplanet case. This can also be understood by comparing Fig. \ref{fig:eqsingle} and Fig \ref{fig:disksingle}.   

In the multi-protoplanet case (Figure \ref{fig:Lparam}b), the giant planet formation zone gets extended to lower-mass and less turbulent disks.  This is because, first, disk mass decreases slowly in disks with long lifetimes, leading to a protracted supply of gas and pebbles. Second, planets with a higher $M_{\rm opt}$ undergo more pronounced convergent migration (Fig. \ref{fig:disksingle}d),  enhancing the probability of planet-planet collisions. Third, $M_{\rm iso}$ is higher at later times. All these factors facilitate the giant planet's  growth in disks with long disk lifetime. 

Therefore, the longer dissipation timescale of turbulent disks promotes the formation of massive solid planetary cores and the accumulation of substantial gas envelopes, resulting in the formation of gas giant planets with a mass approximately 0.7 times that of Jupiter. While in low-turbulence disks, only super-Earths with a few Earth masses can be formed.

\subsubsection{Mass of protoplanets} \label{sec:psmass}

\begin{figure*}[htb]
    \centering
    \includegraphics[width=\linewidth]{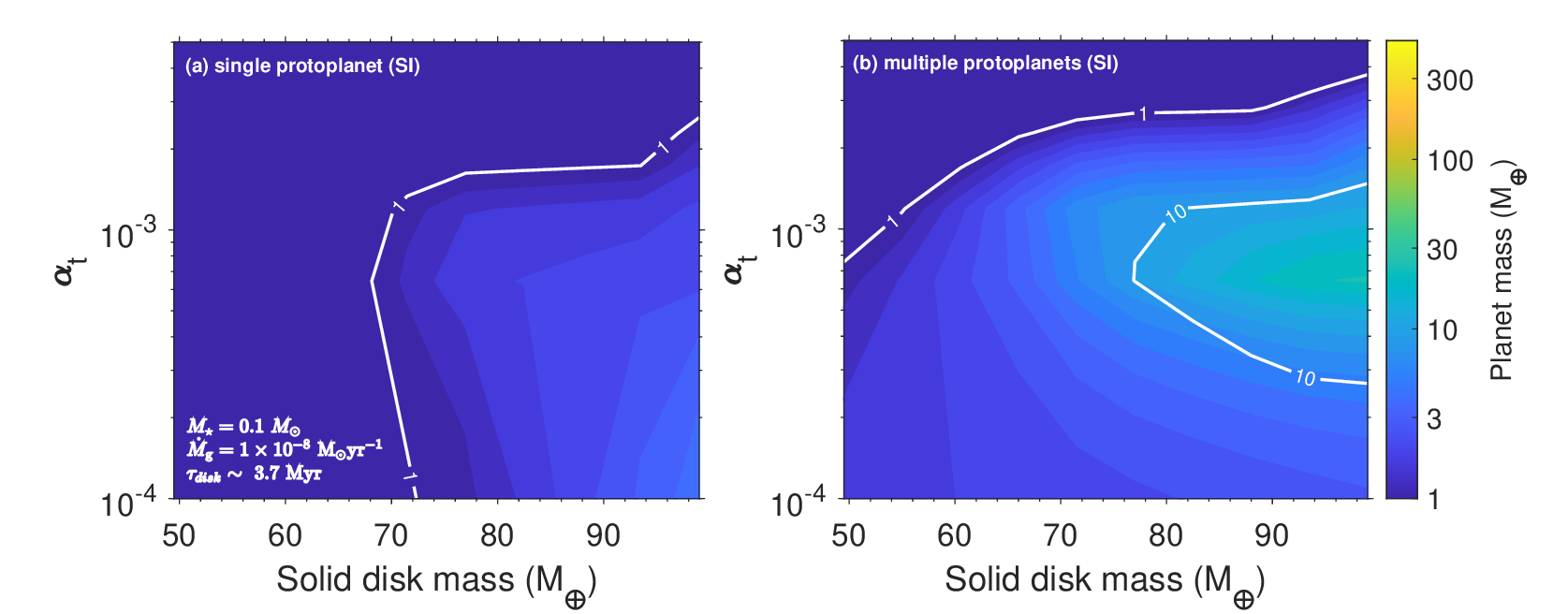}
    \caption{Formation of massive planets in the streaming instability protoplanet mass scenario as a function of disk turbulent level and solid disk mass. The left and right panels illustrate the single protoplanet and mult-protoplanet cases. The protoplanets are assumed to form by streaming instability where their masses follow \Eq{Mp0} and the disk lifetime is $3.7$ Myr. Other model parameters are listed in  Table \ref{tab:param}.  The colorbar gives the resulting planet with the highest mass, while the contour lines indicate masses of $10$, $30$, and $100$, respectively. Giant planets around stars of $M_{\star}{=}0.1 \ M_{\odot}$ are difficult to form when the protoplanets from by streaming instability.  }
    \label{fig:SIparam}
\end{figure*}

We explore the influence of $M_{\rm p0}$ by assuming that protoplanets form by streaming instability (equation \ref{eq:Mp0}). The parameter map is given in  Figure \ref{fig:SIparam}. 

In single protoplanet cases,  planet growth is most efficient at the lowest $\alpha_{\rm t}$ and highest pebble disk mass. This trend is also shown in \Fg{SIsingle}. Since $M_{\rm p0}$ is much lower than lunar mass in most region of the planetary disk, the planets take a longer time to grow their core masses, resulting in final planets with lower masses compared to those start with equal lunar mass in Figure \ref{fig:eqparam}.  

In the case of multiple protoplanets, the optimal zone for massive planet formation shifts to $\alpha_{\rm t}{\sim} 10^{-3}$. This is because protoplanets with shorter orbital distances can grow beyond a few Earth masses and migrate quickly into the inner disk region (\Fg{SIsingle}e). They accumulate in a compact configuration, leading to late-phase giant impacts. It is important to note that by this stage, the disk gas has been substantially dissipated, leaving planets with limited gas envelopes to accrete. Thus, Neptune-mass planets can form in moderately turbulent disks with $M_{\rm solid} \gtrsim 85 \ {\rm M}_{\oplus}$.

\section{Discussion}
\label{sec:discussion}
We discuss the tension between the observed low dust masses and the model required high solid disk masses for giant planet formation in Sect.\ref{sec:dust}. A few aspects of our model are also assessed, including varying Stokes number with disk turbulence, different disk structures, stellar luminosity, and stellar masses in Sect. \ref{sec:stokes}, Sect. \ref{sec:disk},  Sect. \ref{sec:luminosity} and Sect. \ref{sec:starmass}. The limitations and caveats are discussed in Sect. \ref{sec:caveat}.

\subsection{Disk mass budget for planet formation}
\label{sec:dust}

Recent observations of protoplanetary disks in various star-forming regions have shown that their dust masses typically range from a few hundred to a few Earth masses with a huge scatter \citep{Pascucci2016, Long2018, Tobin2020, Tychoniec2020, Miotello2022, Manara2023}. These observations indicate that the solid mass of the disk decreases over time, with the highest values in Class 0 disks and a gradual depletion towards the Class II and III phases (e.g., see Figure 2 in \cite{Drazkowska2023}).  Furthermore, dust mass is lower around M-dwarfs compared to their solar-mass counterparts \cite{Andrews2013, Pascucci2016}. For instance, the average solid mass is only ${\sim} 1 \ M_{\oplus}$ in ${\sim}2$ Myr old Lupus disks around stars of $0.1 \ M_{\odot}$, probably due to the radial drift of pebbles \citep{Appelgren2023}. This poses a serious challenge for the formation of giant planets, as it raises the question of whether such disks contain enough solids to form sufficiently massive planetary cores.

It is worth noting that in previous studies, in order to derive the solid disk mass from dust continuum measurements, two assumptions were made: the dust emission is optically thin, and the opacity is mainly due to absorption rather than scattering. However, both of these assumptions have been called into question. \cite{Zhu2019} and \cite{Liu2019HB} pointed out that optically thick disks with scattering can be misinterpreted as optically thin disks, leading to an underestimation of the disk mass in the literature.

In a recent study by \cite{Macias2021}, both scattering and absorption in dust opacity are considered, without making any underlying assumptions on the optical depth. The authors found that in the TW Hydrae disk, the dust mass is  ${\sim}300 \ M_{\oplus}$, a factor of $5$ or higher than what would be estimated using typical assumptions. Similar findings are also obtained for the study of the disk of low-mass star ZZ Tau IRS \citep{Hashimoto2022}.  These results highlight the importance of considering more realistic dust opacity models in estimating the mass of protoplanetary disks.

On the other hand, the occurrence rate of giant planets around early M dwarfs has been estimated to be less than $5\%$ \citep{Bonfils2013, Sabotta2021}, and an even lower occurrence rate is anticipated around stars of $0.1 \ M_{\odot}$. Therefore, the disk conditions that are preferred for the growth of giant planets cannot be considered typical, but rather represent outliers. Based on the above discussions, it is still likely that early protoplanetary disks around such low-mass stars could contain pebbles with a total mass of ${>} 50 \ M_{\oplus}$.

\subsection{Turbulence-induced fragmentation-limited Stokes number}
\label{sec:stokes}

In the previous sections, we assume pebbles with a constant Stokes number. Nevertheless, disk turbulence could raise the relative motion between solid particles. When the pebbles' relative velocity is dominated by turbulence, their maximum Stokes number can be expressed as $\tau_{\rm s} {\approx}\frac{v_{\rm F}^2}{\alpha_t c_{\rm s}^2}$ \citep{Birnstiel2012}, where $c_{\rm s}$ is the gas sound speed, $v_{\rm F}$ is the fragmentation threshold velocity \citep{Blum2008}. This means that the Stokes number of the largest pebbles, constrained by the fragmentation limit, decreases as turbulence increases\footnote{Note that even in the fragmentation limited, if  the pebbles' relative velocity is not dominated by turbulent (e.g., by radial drift),  $\tau_{\rm t}$ can be independent of $\alpha_{\rm t}$, see discussions in \cite{Drazkowska2021}.}.  We consider this  $\alpha_{\rm t}$ dependence on $\tau_{\rm s}$ in this subsection and assume  $\tau_{\rm s} {=}0.05 \times (10^{-3}/\alpha_{\rm t}) $, with the adoption of  $v_{\rm F} {=} 7 \rm \ m/s$.

\begin{figure*}
    \centering
    \includegraphics[width=\linewidth]{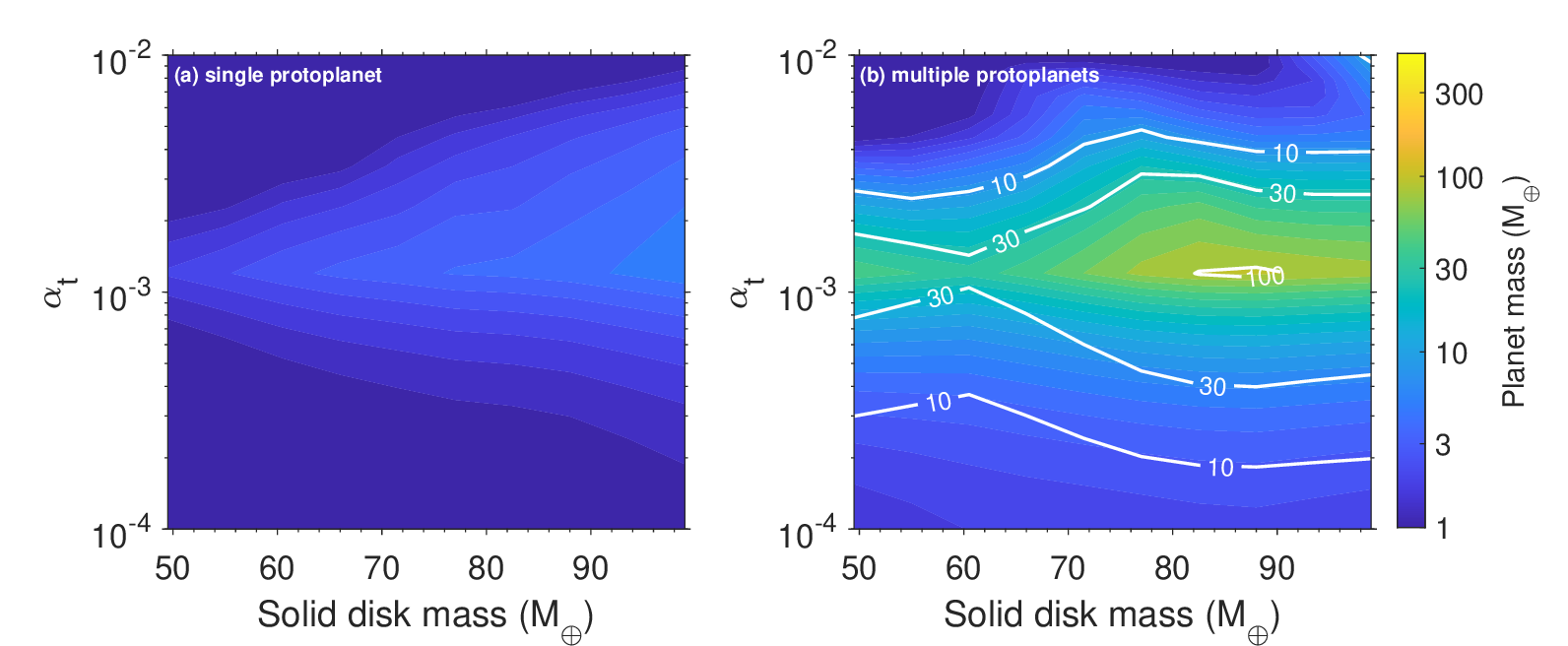}
    \caption{Similar to Fig. \ref{fig:eqparam} but in the fragmentation-limited disk. The stokes number is inversely related to the turbulence strength, and we adopt $\tau_{\rm s} {=} 0.05$ when $\alpha_t {=} 10^{-3}$. The parameters are listed in Table \ref{tab:param2}. The giant planet formation region is significantly narrower compared to the one depicted in Fig. \ref{fig:eqparam}, primarily because the growth of pebble size is suppressed in high-turbulence disks.}
    \label{fig:stokes}
\end{figure*}

We explore this circumstance with parameters presented in Table \ref{tab:param2}, and the result is demonstrated in Fig. \ref{fig:stokes}.   Compared to Fig. \ref{fig:eqparam}, the notable difference occurs for  $\alpha_{\rm t}{\gtrsim} 2 \times 10^{-3}$.  When pebbles' Stokes number is dependent on $\alpha_{\rm t}$,  the growth of pebbles is strongly suppressed in highly turbulent disks. These small particles cannot settle effectively and therefore pebble accretion is largely impeded. As a result, massive giant planets can only form in massive disks with moderately turbulent level ($\alpha_{\rm t}{\sim}10^{-3}$).

\subsection{Pure stellar irradiation disk}
\label{sec:disk}
We test a case with a purely stellar-irradiated disk, keeping all other parameters the same as in Figure \ref{fig:eqsingle}f (see Table \ref{tab:param2}). The results are shown in Figure \ref{fig:irr}a. We observe that the disk's scale height is lower in the inner disk region, leading to the formation of planets with lower pebble isolation mass and rapid inward migration. Super-Earth planets are only favored to form in the outer disk region of $10{-}20$ au. We do not find that giant planets form in disks in the absence of viscously heated regions.

The effect of multiple protoplanets is expected to be limited because there is only inward migration, and inner planets reach lower $M_{\rm iso}$ at earlier times. The migration is rather divergent, so we do not anticipate a significant difference in the final mass of the planet between the multiple protoplanet case and the single protoplanet case. 

\subsection{Stellar luminosity}
\label{sec:luminosity}
We also investigate the impact of stellar luminosity on planet growth. Low-mass young stars typically follow an empirical relation such that $L_{\star} \propto M_{\star}^{\beta}$, where the power-law index $\beta{\sim}1{-}2$. For a star with a mass of $0.1 \ M_{\odot}$, its luminosity typically ranges from $0.01$ to $0.1 \ L_{\odot}$. In our simulations, we adopt a conservative value of $0.01 \ L_{\odot}$ and assume that the stellar luminosity remains constant over the relatively short disk lifetime of several million years.

\begin{figure}
    \centering
    \includegraphics[width=\columnwidth]{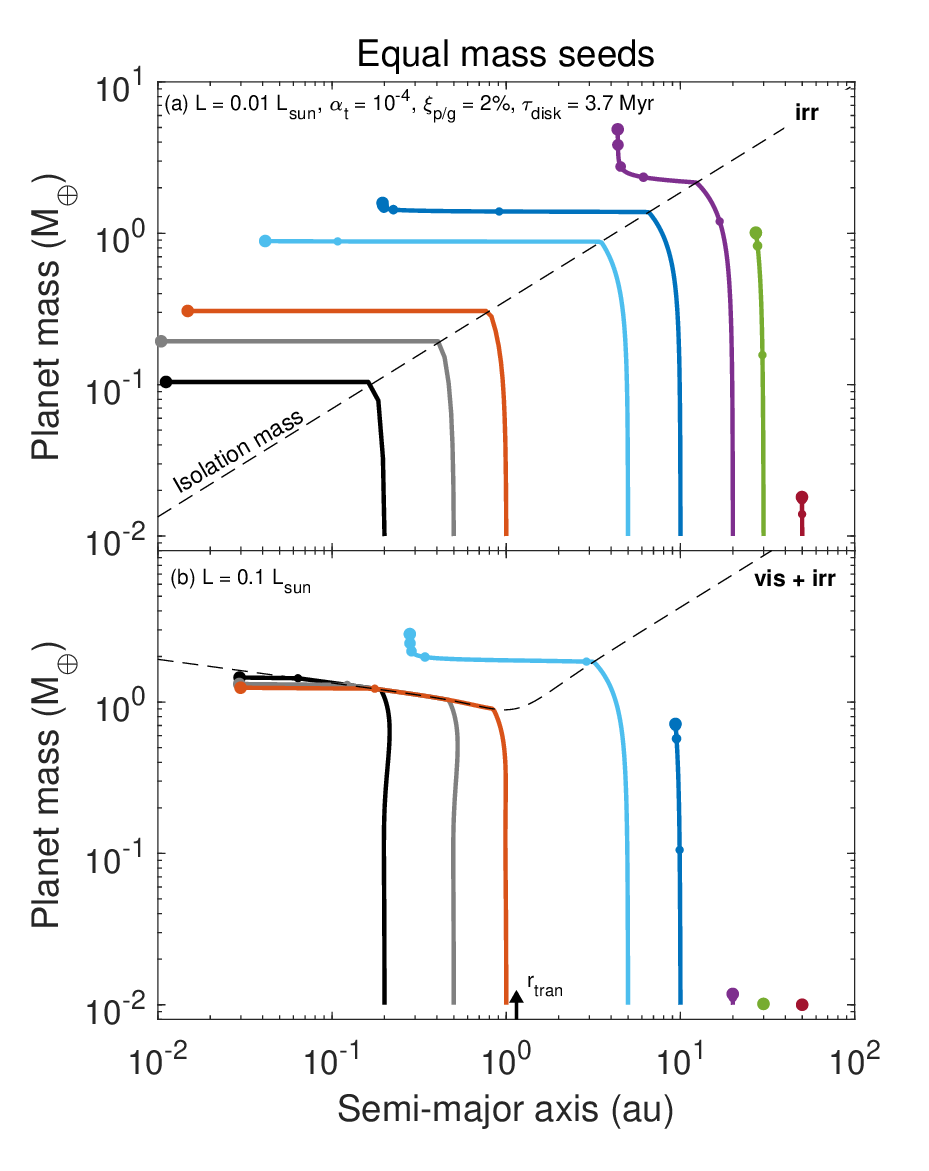}
    \caption{
    Growth and migration of individual protoplanets at different disk locations around stars of $M_{\star}{=}0.1 \ {M}_{\odot}$ in a pure stellar irradiation disk with $L_{\star}{=}0.01 \ L_{\odot}$ (top)  and in a viscously heated and stellar irradiated disk with $L_{\star}{=}0.1 \ L_{\odot}$ (bottom). The model parameters are similar to that in Figure \ref{fig:eqsingle}f (see Table \ref{tab:param2}).
    }
    \label{fig:irr}
\end{figure}

We test a higher value of $0.1 \ L_{\odot}$ while the other model parameters are the same as in Figure \ref{fig:eqsingle}f. Our simulations show that planet growth becomes more difficult in disks around more luminous stars. In this circumstance, the stellar irradiation region becomes more predominant, and the gas disk scale height is much larger, resulting in extremely low pebble accretion rates. Even in the most metal-rich disks we explored ($\xi_{\rm p/g}{=}2\%$), growth remains slow and the protoplanet hardly reaches $M_{\rm iso}$ in the outer disk region. Only super-Earth planets of $2 \ M_{\oplus}$ can form at a moderate $r_0{\sim}5 $ au (Fig. \ref{fig:irr}b). In this regard, we expect that massive planets are preferred to form in the late stage of protostellar evolution.

It is important to note that in reality, the stellar luminosity should gradually decline over time \citep{Chabrier1997, Baraffe2015}, and the more realistic planet growth pattern lies somewhere between the cases of constant low and high luminosity (Fig. \ref{fig:eqsingle}f and Fig. \ref{fig:irr}b).  In future work,
we intend to investigate planet formation coupled with a more self-consistent time-dependent evolution of stellar luminosity.

\begin{figure*}
    \centering
    \includegraphics[width=\linewidth]{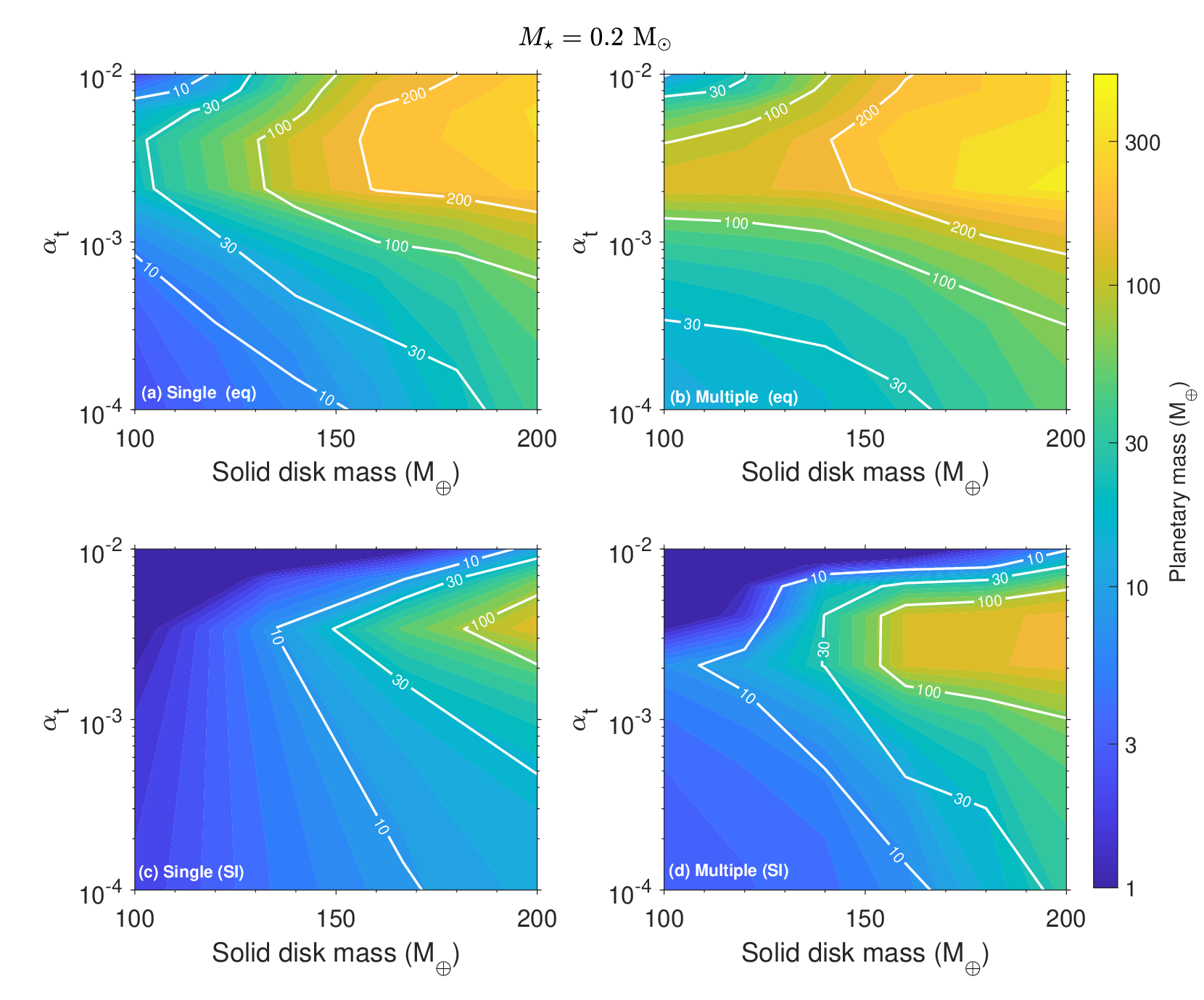}
    \caption{Similar to Fig. \ref{fig:eqparam} (panel a and b) and Fig. \ref{fig:SIparam} (panel c and d) but for the stellar mass of $0.2 \ {\rm M}_{\odot}$. The disk accretion rate and stellar luminosity increase with stellar following $\dot{M}_{\star} \propto M_{\star}$ and $L_{\star} \propto M_{\star}^2$, respectively. Larger gaseous planets are achieved around more massive stars even in the case that protoplanets form through streaming instability.
    }
    \label{fig:S2param}
\end{figure*}

%\subsection{Distant giant planets}

\subsection{Stellar mass}
\label{sec:starmass}

In addition to the influence of stellar luminosity on the disk profile, the stellar mass is an important factor affecting the availability of supplementary materials within the protoplanetary disk. This, in turn, governs the amount of solid material that protoplanets can accrete.
Here we explore the growth and evolution of protoplanets around stars of $0.2 \ M_{\odot}$ by assuming a linear scaling relationship between the disk mass and stellar mass. Figure \ref{fig:S2param} displays the maximum planetary mass attained by a single protoplanet as well as multiple protoplanets formed with a lunar mass or through streaming instability.

In the case of a single protoplanet with equal mass, the formation of gaseous planets with mass $> 30 \ M_{\oplus}$ is possible in massive disks with $M_{\rm d} \geq 150 \ M_{\oplus}$ or lower mass disks whose $\alpha_{\rm t} \gtrsim 10^{-3}$. The giant planet formation zone extends significantly beyond the fiducial case around $0.1 \ M_{\oplus}$ (see Fig. \ref{fig:eqparam}a) for three reasons. Firstly, the pebble isolation mass increases with the stellar mass (Eq. \ref{eq:Miso}). Protoplanets orbiting more massive stars can therefore achieve a higher core mass and accumulate a denser atmospheric envelope. Secondly, both the inward migration timescale (Eq. \ref{eq:tim}) and the optimal mass for outward migration (Eq. \ref{eq:opt}) increase with the stellar mass, favoring the growth of protoplanets in the outer regions of the disk. Lastly, the pebble and gas fluxes are also enhanced with stellar mass, providing a greater supply of materials for protoplanet growth.

When considering the mutual interactions of multiple equal-mass protoplanets (Fig. \ref{fig:eqparam}b), the giant planets are widely available for $\alpha_{\rm t} \geq 5 \times 10^{-4}$ due to planet-planet collisions. Whereas the formation for gas giants larger than $0.5 \ M_{\rm J}$ is still limited in the turbulent disk with $\alpha_{\rm t} > 10^{-3}$ and the solid disk mass $> 150 \ M_{\odot}$.

Another significant distinction in planet formation around stellar masses of $0.1 \ M_{\odot}$ and $0.2 \ M_{\odot}$ is that giant planets can form from seeds generated by streaming instability (Fig. \ref{fig:eqparam}c and d). Gas giants with masses greater than $100 \ M_{\oplus}$ are exclusively formed in single-protoplanet scenarios when the disk has a turbulent viscosity parameter $\alpha_{\rm t} = 5 \times 10^{-3}$, accompanied by a solid disk mass of approximately $200 \ M_{\odot}$. In the case of multiple protoplanets, the gas giant formation zone expands to $10^{-3} \lesssim \alpha_{\rm t} \lesssim 6 \times 10^{-3} $ and a solid disk mass exceeding $150 \ M_{\odot}$.

\subsection{Comparison with Observations}

We have demonstrated the possibility of forming gas giants with masses ranging from 0.1 to 0.4 $M_{\rm J}$ around stars with a mass of $0.1 \ M_{\odot}$, which is consistent with the discovery of four giant planets orbiting host stars with $M_{\star} < 0.2 \ M_{\odot}$, namely TOI-1227 b ($ <0.5 \ M_{\rm J}$), GJ 3512 b ($0.46 \ M_{\rm J}$), GJ 3512 c ($0.45 \ M_{\rm J}$), and GJ 9066 c ($0.21 \ M_{\rm J}$). We also provide an illustrative example of a giant planet formation ($0.3 \ M_{\rm J}$) promoted by pebble accretion and planet-planet collisions at a location of approximately 0.09 au (see Fig. \ref{fig:eqmulti}), exhibiting an agreement with the transit observation of planet TOI-1227 b.

However, we admit that the smooth disk assumption in our study faces challenges in explaining the orbital characteristics of the other three distant giant planets discovered by radial velocity surveys. The rapid inward migration of planets, predominantly caused by the Lindblad torque, occurs prior to the formation of a surface density gap \citep{Paardekooper2011}.
We anticipate that a structured disk with rings and gaps would effectively suppress the inward migration of planets \citep{Baillie2016} and facilitate the formation of giant planets in wide orbits. We plan to explore this possibility in our future work.
Besides, the interactions between the planets and the gaseous disk result in orbital circularization \citep{Kley2012}. In order to reproduce the observed high orbital eccentricities of GJ 9066 c, it is essential to consider the close encounters and mutual scatterings during the later dynamical evolution of planets in a gas-free environment \citep{Ji2011, Ida2013}.

\subsection{Caveat}
\label{sec:caveat}

\cite{Bell1994}'s opacity law is widely adopted in the community, which is based on the assumption of ISM-like grains with compact spherical structures. However, various physical processes have taken place in protoplanetary disk environments (e.g., grain growth and crystallization), and the corresponding disk dust opacity could significantly differ from the ISM's form.   A realistic opacity calculation relies on the detailed size distribution, composition and porosity of the dust population, which yet remains poorly understood.  In this study we assume a simple opacity law. The opacity variation across the ice lines is also neglected. It is worth pointing out that the migration direction might be reversed and planets get trapped at distinctive regions due to opacity transition \citep{Kretke2012}. In our study there is only one convergent migration radius. When considering multiple transition radii,  although the individual growth pattern differs, the optimal disk condition obtained in this work (e.g., disk mass) for giant planet formation still generally holds.

Besides, we adopt a simplified approach that assumes a constant Stokes number of pebbles and a fixed pebble-to-gas flux ratio under all varied disk and stellar environments (except in Section \ref{sec:stokes}). The assumption of a fixed pebble-to-gas flux ratio can be justified if pebbles remain small in the outermost regions of the protoplanetary disc \citep{Johansen2019}, so that the pebble flux follows closely the gas mass flux. In this approach, pebbles may still grow large in the inner region of the disc, even though their total flux remains small and hence the pebble population is not depleted.  In more realistic conditions, the coagulation and fragmentation of dust result in them following a power-law size distribution \citep{Birnstiel2011}, and the pebble flux is not necessarily always attached to gas flow \citep{Drazkowska2023}. In future work, we aim to implement a more sophisticated dust-size population and flux profile (e.g., from Dustpy code, \cite{Stammler2022}) to gain a better understanding of how these factors influence final planet growth.

We also utilize a simplified gas accretion model that disregards the influence of disk temperature on the gas accretion rate. It has been demonstrated that the elevated temperatures closer to the central star result in higher thermal energy, posing a challenge for the hot gas to be gravitationally captured by and accreted onto the planet \citep{Coleman2017}. Planets located within the innermost disk region require a more extended period to cool down, leading to a reduction in the gas accretion rate \citep{Piso2014}. According to \cite{Lee2014}), the runaway accretion timescale follows a power-law relationship with disk temperature ($\tau_{run} \propto T^{0.34}$). In such circumstances, the giant planet form at a relatively later stage when the inner gaseous disk has cooled slightly but the disk surface density remains high.
We also note that in our simulations, some planets initiate gas accretion beyond the water snowline and subsequent migrate inward. These planets accrete a substantial amount of gas before entering the innermost disk region.

We note that in Section \ref{sec:luminosity} we have demonstrated that the stellar luminosity influences the disk thermal structure and therefore pebble accretion mass growth. The stellar luminosity cannot remain constant during its whole evolution. In particular, in the early stage, the stars are more luminous and pebble accretion may hardly be effective. The core growth can proceed when stars gradually cool. Since we mainly consider the stars with relatively low luminosities, the condition obtained in this paper can be treated as an optimistic perspective. We leave the implementation of a proper stellar evolution in our following studies.

\section{Conclusions}
\label{sec:conclusion}

In this paper we investigate the formation of giant planets around late M dwarf stars with a stellar mass range between $0.1$ and $0.2 \ M_{\odot}$. Although these planets are rare in exoplanet surveys, their formation mechanism remains unclear. Previous studies have suggested that the core accretion scenario faces difficulties in explaining the existence of such high planet-to-star mass ratio systems around these small stars \citep{Liu2019a, Coleman2019, Liu2020b, Miguel2020, Burn2021}.

To address the issue of whether these giant planets can form through core accretion scenario, we use the pebble-driven planet formation model proposed by \cite{Liu2019a} and perform N-body simulations to study the growth and migration of single and multiple protoplanets in the protoplanetary disk with inner viscously heated and outer stellar irradiated regions. Our simulations incorporate various physical processes, including pebble accretion onto planet cores, gas accretion onto planet envelopes, planet-planet interactions/collisions, type I and type II planet migration and gas damping. We study the influence of several key disk and planet properties, such as the disk turbulent level, solid disk mass (or flux ratio of pebbles and gas), disk lifetime, birth mass of the protoplanets and stellar mass.

The most favorable region for planet growth is near the transition radius $r_{\rm tran}{\sim}3$ au that separates the inner viscously heated and outer stellar irradiated regions. However, in the case of single protoplanet growth, it is difficult for the planet to accrete a massive gaseous atmosphere due to its low pebble isolation mass, which is typically around ${\sim}2{-}3 \ M_{\oplus}$ in systems around stars of  $M_{\star}{=} 0.1 \ M_{\odot}$ (\Fg{eqsingle1}). When considering multi-protoplanets with the same disk condition, their core growth is no longer limited by pebble isolation. Planets massive enough can undergo convergent migration and evolve into tightly compact orbits, which likely induces subsequent orbital crossings and planet-planet collisions. In general, this dynamical process can overcome the pebble isolation mass barrier for the single protoplanet and promote the growth of a massive core even in very low-mass stellar host systems (\Fg{eqmulti}).

Two different birth masses of protoplanets are considered. On the one hand we consider the protoplanets form from runaway/oligarchic planetesimal accretion and end up with masses of $0.01 \ M_{\oplus}$. 
The parameter space for giant planet formation significantly expands when we take into account the growth of multiple protoplanets. Gaseous planets with masses exceeding $100 \ M_{\oplus}$ can form around stars with mass of $0.1 \ M_{\odot}$ in disks characterized by $\alpha_{\rm t} {>} 10^{-3}$ and solid mass ${\gtrsim} 60 \ {\rm M}_{\oplus}$ (\Fg{eqparam}). 
More massive planets are preferred to grow in disks with a longer lifetime and higher supply of pebble reservoirs (\Fgs{disksingle}{Lparam}). Meanwhile, the giant planet formation benefits from the increasing stellar mass, due to high pebble isolation mass, massive solid disks and long planet migration timescale in systems around massive stars.

On the other hand, in the streaming instability scenario the birth mass of protoplanets increases with orbital distance and stellar mass.   Generally, $M_{\rm p0}$ is much lower than $0.01 \ M_{\oplus}$ at $r{<} 10$ au. In the single protoplanet case, super-Earth planets only form in the outer region of low-turbulence disks around $0.1 \ M_{\odot}$ stars (\Fg{SIsingle}). Neptune-mass planets can form in the multiple protoplanet case in disks with $\alpha_{\rm t} {\sim} 10^{-3} $ and solid mass exceeding $80 \ M_{\oplus}$ (\Fg{SIparam}). For systems around more massive stars of $0.2 \ M_{\odot}$, the formation of giant planets takes place in disks with $10^{-3} {\lesssim} \alpha_{\rm t} {\lesssim} 6 \times 10^{-3} $ and solid disk masses ${>} 150 \ M_{\odot}$ (\Fg{S2param}).

Overall, our study highlights the crucial finding that the formation of giant planets with orbital periods of  ${\lesssim} 100$ days is favored in turbulent and massive protoplanetary disks. The extended lifetime of the disk and a higher stellar mass contribute to the formation of more massive planets, despite a narrower formation zone within long-lived disks. If protoplanets arise from streaming instability, they only give rise to the birth of giant planets when the stellar mass exceeds $0.2 \ M_{\odot}$.

We propose the formation of giant planets with masses ranging from $0.1$ to $0.6 \ M_{\rm J}$ around stars with masses of $0.1 \ M_{\odot}$. This finding aligns with the observed planetary mass in GJ 3512, GJ 9066, and TOI-1227 systems, which were studied through the CARMENES and TESS programs \citep{Morales2019, Mann2022, Quirrenbach2022}. Furthermore, our results suggest an increasing feasibility of giant planet formation as stellar mass increases, indicating a correlation between the occurrence rate of giant planets and stellar mass \citep{Bryant2023, Gan2023b, Ribas2023}.
We anticipate that ongoing and upcoming exoplanet search projects, such as TESS \citep{Ricker2015}, MEarth \citep{Irwin2009}, TRAPPIST \citep{Jehin2011}, SPECULOOS \citep{Sebastian2021}, CARMENES \citep{Quirrenbach2014}, EDEN \citep{Gibbs2020}, PLATO \citep{Rauer2014}, ET \citep{Ge2022}, and CHES \citep{Ji2022}, will provide a larger sample of planets with well-constrained mass and orbital properties. 
As the gas accretion at different place indicates a different gas composition of giant planets, we also expect that JWST observations offer valuable insights into the composition of planetary atmospheres to constrain the birthplaces of giant planets.
These datasets will significantly contribute to our understanding of planetary formation around very low-mass stars.

\begin{acknowledgements}
    We thank Xuening Bai, Zhaohuan Zhu, Gillon Micha\"{e}l, Amaury Triaud, Haifeng Yang for useful discussions. We also thank the anonymous referee for their useful suggestions and comments. BL and MP are supported by National Natural Science Foundation of China (Nos. 12222303, 12173035 and 12111530175), the start-up grant of the Bairen program from Zhejiang University and the Fundamental Research Funds for the Central Universities (2022-KYY-506107- 0001,226-2022-00216). 
    A.J. acknowledges funding from the European Research Foundation (ERC Consolidator Grant 724687-PLANETESYS), the Knut and Alice Wallenberg Foundation (Wallenberg Scholar Grant 2019.0442), the Swedish Research Council (Project Grant 2018-04867), the Danish National Research Foundation (DNRF Chair Grant DNRF159) and the G\"{o}ran Gustafsson Foundation. 
    W.S. is funded by the National Natural Science Foundation of China (Nos. 12033010, 12111530175), the B-type Strategic Priority Program of the Chinese Academy of Sciences (Grant No. XDB41000000), Foundation of Minor Planets of the Purple Mountain Observatory.
    J.J appreciate support from the National Natural Science Foundation of China (Grant Nos. 12033010), the B-type Strategic Priority Program of the Chinese Academy of Sciences (Grant No. XDB41000000), Foundation of Minor Planets of the Purple Mountain Observatory.
    I.R. acknowledges financial support from the Agencia Estatal de Investigaci\'on of the Spanish Ministerio de Ciencia e Innovaci\'on MCIN/AEI/10.13039/501100011033 and the ERDF ``A way of making Europe'' through project PID2021-125627OB-C31, from the Centre of Excellence ``Mar\'{\i}a de Maeztu'' award to the Institut de Ci\`encies de l’Espai (CEX2020-001058-M) and from the Generalitat de Catalunya/CERCA programme.
    The computations are supported by cosmology simulation database (CSD) in the National Basic Science Data Center (NBSDC-DB-10).
\end{acknowledgements}


\begin{appendix}

\section{Isolation mass} \label{app:iso}

To provide a clearer explanation of the pebble isolation mass used in this study (as depicted in \Eq{Miso}), we compare it with the pebble isolation masses proposed by \cite{Bitsch2018} (orange line) and \cite{Ataiee2018} (yellow line) in \Fg{iso}. This comparison is specifically conducted for a Stokes number of $\tau_{\rm s} = 0.05$ and an aspect ratio of $h_{\rm g} = 0.05$.

Differ from the 3D hydrodynamical simulations performed by \cite{Bitsch2018}, 
\cite{Ataiee2018} conducted 2D gas hydrodynamical simulations to investigate the minimum planet mass required to create a radial pressure bump beyond the planet's orbit as a function of the disk aspect ratio ($h_g$), the turbulent viscosity ($\alpha_{\rm t}$). Successful particle trapping are further performed by 2D gas plus dust hydrodynamical simulations to explore the effects of dust turbulent diffusion on particle trapping at the pressure maximum.

Both \cite{Ataiee2018} and \cite{Bitsch2018} explored how the local disk parameters influence the pebble-isolation mass. In two-component disk model, the results exhibit significant variation in the low-$\alpha_{\rm t}$ regime, attributed to a harder gap formation in the 3D disk model than in 2D model. Even though, they are still not sufficient to promote runaway gas accretion to form giant planets. In high-turbulent disks, \cite{Ataiee2018} and \cite{Bitsch2018} have reported comparable isolation masses of approximately $10 \ M_{\oplus}$, which is about 2.5 times larger than that in our work. If we adopt the higher isolation masses, it indeed pose a greater challenge for planets to reach isolation and initiate runaway gas accretion. If it were achievable, it would likely occur at a later stage, when the isolation mass has decreased to a lower value or when the disk has largely dissipated. Consequently, the formation of giant planets in turbulent disks is less likely compared to the formation scenario presented in our model.

Furthermore, the gap opening mass should be larger than the pebble isolation mass to account for a deeper gap. However, the results of both \cite{Ataiee2018} and \cite{Bitsch2018} conflict with the value of $M_{\rm iso}$ derived from the 2D hydrodynamical simulations conducted by \cite{Kanagawa2015} when $\alpha_{\rm t} \lesssim 6 \times 10^{-4}$. Thus, we adopt the relationship between $M_{\rm gap}$ and $M_{\rm iso}$ given by \cite{Johansen2019}, where $M_{\rm iso}$ is approximately $2.3$ times smaller than $M_{\rm gap}$. The pebble isolation mass we used is about half of what was reported in \cite{Bitsch2018}, but it exhibits a consistent decrease with lower turbulent viscosity $\alpha_{\rm t} \lesssim 10^{-3}$.

\begin{figure}
    \centering
    \includegraphics[width=\columnwidth]{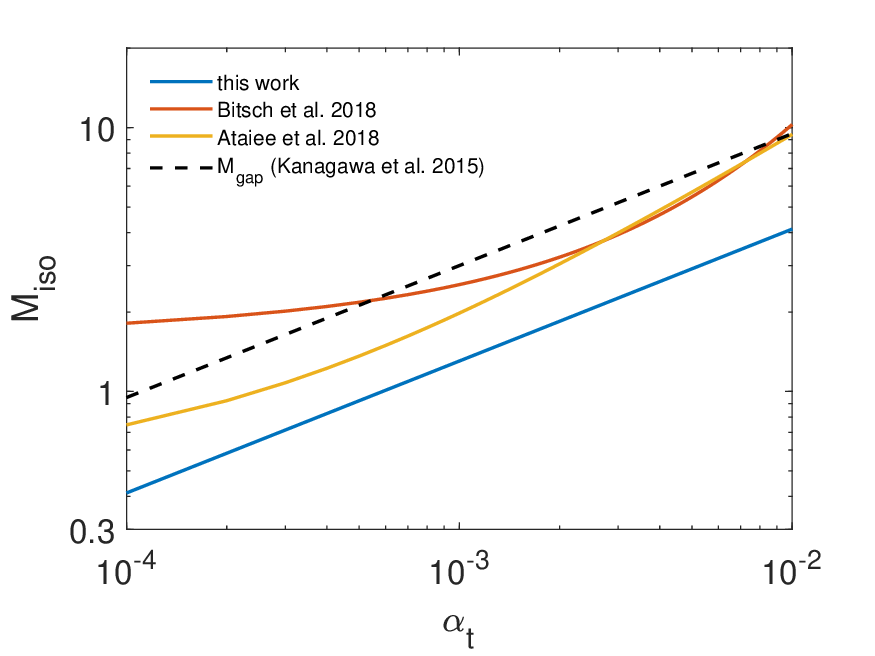}
    \caption{Pebble isolation mass as a function of $\alpha_{\rm t}$ when Stokes number $\tau_{\rm s}{=}0.05$ and aspect ratio $h_{\rm g}{=}0.05$. The results of \cite{Ataiee2018}'s 2D hydrodynamical simulations and the 3D simulations of \cite{Bitsch2018} are shown in yellow and orange lines, respectively. The blue line represents the prescription we used in \Eq{Miso}. The gap openging mass derived from \cite{Kanagawa2015} is indicated by dashed line.
    }
    \label{fig:iso}
\end{figure}

\begin{figure*}
    \centering
    \includegraphics[width=\linewidth]{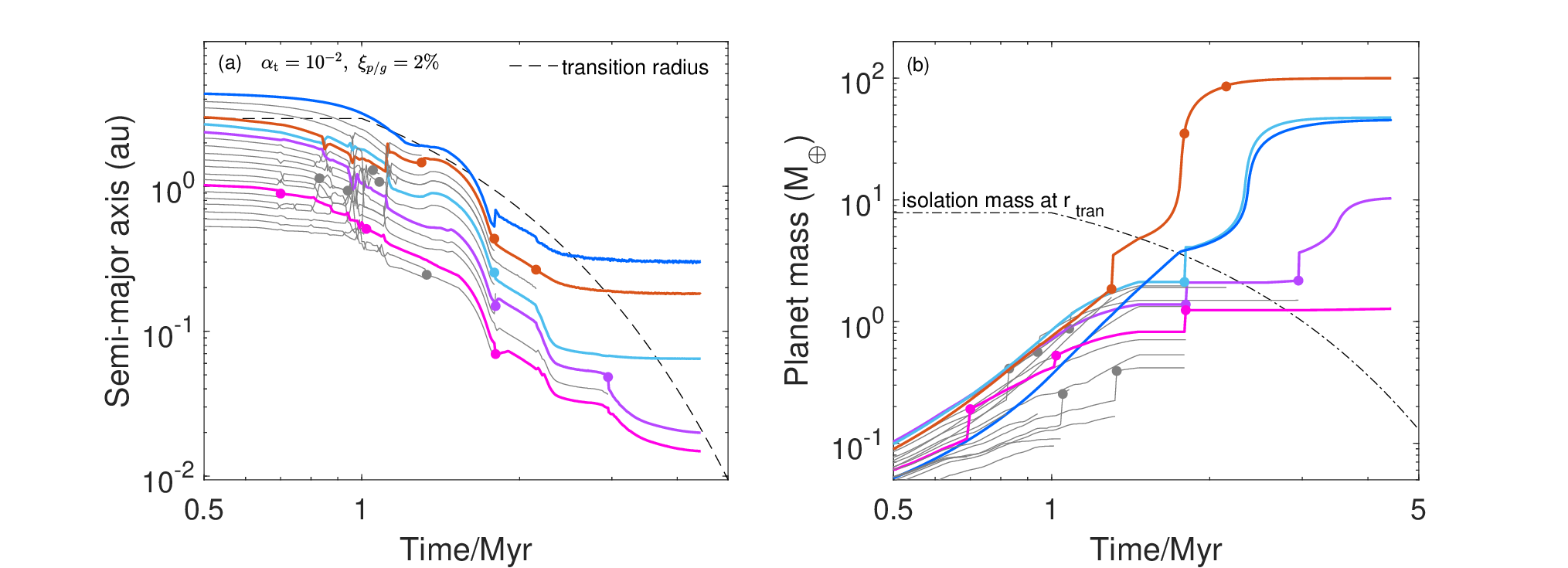}
    \caption{Similar to Fig. \ref{fig:eqmulti}, but for the disk parameters of $\alpha_{\rm t} {=} 10^{-2}$, $\xi_{\rm p/g}{=}2\%$. After stochastic collisions between protoplanets in this high-turbulent disk, the protoplanet represented by the orange solid line undergo rapid growth and trigger gas accretion at approximately 1.3 Myr. It subsequently evolves into a gas giant with an orbital period of 100 days.}
    \label{fig:eqmulti2}
\end{figure*}

\begin{table*}
\caption{List of notations.}
\centering
\begin{tabular}{l l} 
    \hline
    \hline
    Symbol & Description \\
    \hline
    $\alpha_{\rm g}$ & Global angular momentum transfer efficiency \\
    $\alpha_{\rm t}$ & Local turbulent viscosity \\    
    $\eta$ & Headwind prefactor \\    
    $f_{\rm tot}$ & Migration coefficient \\
    $f_{\rm I}$ & Type I migration coefficient \\
    $f_{\rm II}$ & Type II migration coefficient \\
    $f_{\rm s}$ & Smooth function \\
    $\gamma$ & Dimensionless gravity parameter \\
    $\Gamma$ & Total migration torque \\
    $\Gamma_{\rm 0}$ & Normalized migration torque \\      
    $L_{\star}$ & Stellar luminosity \\
    $M_{\odot}$ & Stellar mass \\
    $M_{\rm p}$ & Planet mass \\
    $M_{\rm p0}$ & Initial mass of protoplanet \\
    $M_{\rm opt}$ & Optimal planet mass for outward migration \\
    $M_{\rm iso}$ & Pebble isolation mass \\
    $M_{\rm gap}$ & Gap opening mass \\
    $\dot{M}_{\rm g}$ & Gas disk accretion rate \\
    $\dot{M}_{\rm peb}$ & Pebble mass flux \\ 
    $\dot{M}_{\rm PA}$ & Pebble mass accretion rate onto the planet \\
    $\dot{M}_{\rm p,g}$ & Gas accretion rate onto planet \\
    $h_{\rm g}$ & Gas disk aspect ratio \\
    $h_{\rm g,vis}$ & Gas disk aspect ratio in viscously heated region \\
    $h_{\rm g,irr}$ & Gas disk aspect ratio in stellar irradiated region \\
    $h_{\rm peb}$ & Pebble disk aspect ratio \\
    $\kappa$ & Disk opacity \\
    $\kappa_{\rm 0}$ & Disk opacity coefficient \\
    $\kappa_{\rm env}$ & Envelope opacity of planet\\  
    $r$ & Distance between the planet and the central star \\
    $r_{\rm tran}$ & Transition radius between viscously heated and stellar irradiated regions \\
    $\Sigma_{\rm g,vis}$ & Gas surface density in viscously heated region \\
    $\Sigma_{\rm g,irr}$ & Gas surface density in stellar irradiated region \\
    $T_{\rm g}$ & Gas disk temperature \\
    $T_{\rm g,vis}$ & Gas disk temperature in viscously heated region \\
    $T_{\rm g,irr}$ & Gas disk temperature in stellar irradiated region \\
    $\tau_s$ & Pebble Stokes number \\
    $t_{\rm 0}$ & Onset time of disk dissipation \\
    $\tau_{\rm dep}$ & Disk dissipation timescale \\
    $t_{\rm disk}$ & Timespan for the gas surface density at $1$ au drops to $1 \ {\rm g} \, {\rm cm}^{-2}$ \\
    $\varepsilon_{\rm PA}$ & Totoal pebble accretion efficiency \\
    $\varepsilon_{\rm PA,2D}$ & 2D pebble accretion efficiency \\
    $\varepsilon_{\rm PA,3D}$ & 3D pebble accretion efficiency \\
    $v_{\rm K}$ & Keplerian velocity at planet’s location \\
    $v_{\rm esc}$ & Escape velocity of the planetary system \\
    $\Delta v$ & Relative velocity between the pebbles and planet \\
    $\xi_{\rm p/g}$ & Pebble-to-gas mass flux ratio \\  
    \hline
\end{tabular}    
\end{table*}
\end{appendix}

\end{document}